\documentclass[preprint,10pt]{elsarticle} 
\makeatletter
\def\ps@pprintTitle{%
	\let\@oddhead\@empty
	\let\@evenhead\@empty
	\def\@oddfoot{}%
	\let\@evenfoot\@oddfoot}
\makeatother
\usepackage[left=15mm,right = 15mm,top = 15mm,bottom = 15mm]{geometry}
\usepackage[titletoc]{appendix}
\usepackage[table]{xcolor}
\usepackage{float}
\usepackage{subcaption}
\usepackage{amssymb,amsmath,amsthm,graphics,psfrag,graphicx,color,fancyhdr}
\usepackage{algorithmic}
\usepackage[color=yellow]{todonotes}

\usepackage{appendix}
\usepackage{verbatim}
\usepackage{blindtext}
\usepackage{hyperref}
\usepackage{cleveref}
\usepackage{stmaryrd}
\usepackage{soul}
\usepackage{sidecap}
\usepackage{tikz}
\usepackage{rotating}
\usepackage{booktabs}
\usepackage{comment}
\usepackage{multirow}
\usepackage{changepage}
\usepackage{float}
\usepackage{wrapfig}
\usepackage{stmaryrd}
\usepackage{pgfplots}
\usepackage{framed}
\usepackage{float}

\usepgfplotslibrary{fillbetween}

\crefformat{appendix}{#2#1#3}

\definecolor{lightblue}{rgb}{0.32,0.45,0.90}
\definecolor{lightgreen}{rgb}{0.42,0.7,0.40}
\numberwithin{equation}{section}
\numberwithin{figure}{section}
\hypersetup{colorlinks=true,linkcolor = blue,citecolor=blue}
\usetikzlibrary{shapes,arrows}
\usetikzlibrary{decorations.markings}
\numberwithin{figure}{section}

\def\b{\boldsymbol}

\makeatletter
\newcommand{\vast}{\bBigg@{4}}
\newcommand{\Vast}{\bBigg@{5}}
\newcommand{\scp}[2]{\left<#1\,,\,#2\right>}
\makeatother\def\b{\boldsymbol}

\newcommand\rd{\mathrm{d}}

\newcommand\bu{{\boldsymbol{u}}}

\newcommand\bx{\boldsymbol{x}}

\newcommand\D{{\mathrm D}}

\colorlet{cgray}{gray!20!white}

\theoremstyle{definition}

\newtheorem{method}{Method}[section]

\theoremstyle{remark}

\newtheorem{algorithm}{Algorithm}[section]

\tikzset{every label/.style={font=\footnotesize,inner sep=1pt}}

\allowdisplaybreaks

\pgfplotsset{compat=1.18} 

\begin{document}
\begin{frontmatter}
\title{Comparing two different types of stochastic parametrisation in geophysical flow}
\author[inst4]{D.D. Holm}
\author[inst4]{ W. Pan}
\author[inst4]{J.M. Woodfield}

\affiliation[inst4]{organization={Department of Mathematics, Imperial College London},
addressline={South Kensington Campus}, 
city={London},
postcode={SW7 2AZ}, 
country={United Kingdom}}

\begin{abstract}
This paper investigates the effects of stochastic variations in bathymetry on the solutions of the thermal quasigeostrophic (TQG) equations. These stochastic perturbations generate a variety of different types of ensemble spread in the solution behaviour whilst also preserving the deterministic Lie Poisson structure and Casimir conservation laws. We numerically compare the solution sensitivity, to another type of structure-preserving stochastic perturbation where instead of bathymetry, the velocity is stochastically perturbed. 

\end{abstract}
\end{frontmatter}

\section{Introduction}

In using stochastic modelling to estimate uncertainty in geophysical fluid dynamics (GFD) one may want to keep in mind that in GFD flows the potential energy is often much greater in magnitude than the kinetic energy, especially if thermal effects are involved.  Thus, the effects of uncertainty in potential energy could play a significant role in many GFD flows. \smallskip


The SALT (Stochastic Advection by Lie Transport) approach for stochastic fluid dynamics was derived in \cite{holm2015} using a stochastic version of Hamilton's variational principle for ideal fluids. The SALT approach introduces a stochastic transport velocity \cite{holm2021stochastictqg}. Upon Legendre transforming to a stochastic Hamiltonian description, one notices that another option is also available for introducing stochasticity into fluid dynamics. Namely, one may introduce stochasticity in the Hamiltonian framework for fluid dynamics by introducing stochastic perturbations into the Hamiltonian in either the kinetic energy \textit{or} the potential energy. The former choice produces the SALT fluid equations. The latter choice produces the SPEC (Stochastic Potential Energy Coupling) approach which introduces a stochastic force into the fluid motion equation derived from variations of the potential energy of the stochastic fluid Hamiltonian. 

From one viewpoint, the theoretical SPEC approach here is reminiscent of the Stochastically Perturbed Parametrization Tendencies (SPPT) approach for numerical simulations of atmospheric weather \cite{palmer2009stochastic,buizza1999stochastic} framework. In the SPPT simulation framework, stochastic perturbations are inserted into physics modules in an effort to represent model uncertainty.

In this paper, we compare the effects of stochastic perturbation in either the bathymetry (SPEC) or the transport velocity (SALT) on solutions of the Thermal Quasi-Geostrophic (TQG) equations. As it turns out for the cases studied here, TQG solutions are significantly affected by \textit{both} types of stochastic perturbations, SALT and SPEC. 


\subsection{Derivations of deterministic and stochastic TQG}
Derivations of the well-known deterministic TQG equations appear in, e.g., \cite{ripa1996linear,warneford2013quasi, crisan2023atheoretical}. Stochastic TQG equations have recently been derived via an asymptotic expansion of the variational principle for the stochastic Euler-Boussinesq equations \cite{holm2021stochastictqg}. In Crisan et al. \cite{crisan2023btheoretical} the analytical properties of the new stochastic TQG equations were established for periodic boundary conditions. In particular, existence of local-in-time unique strong solutions was proved in the stochastic TQG model and blow-up criteria of TQG were also identified in Crisan et al. \cite{crisan2023btheoretical}.

\subsection{Motivation}



Ocean Bathymetry has been measured using various modern measurement devices, including Single Beam Echo-Sounders, Multibeam Echo-Sounders, as well as inferred from Satellite data. Despite this recent progress, less than 20 percent of the seafloor has been determined (using echo-sounders at a resolution of about 1 km) \cite{mayer2018nippon,wolfl2019seafloor}, and interpolation must often be performed in dealing with the remaining topography of the ocean floor.

In this paper, we will provide evidence that the uncertainty in solutions of the TQG equations depend about as strongly on the potential energy variations in bathymetry as they depend on the kinetic energy variations which determine the transport velocity. Consequently, we conclude that it is important to account for the uncertainty associated with both bathymetry and transport velocity. We will propose a stochastic parametrisation to take into account uncertainty arising from measurement, interpolation and other types of bathymetry uncertainty whilst retaining key aspects of the deterministic model. The stochastic parametrisation we call SPEC (Stochastic Potential Energy Coupling) plays a significant role in the TQG solution sensitivity which is comparable to the TQG sensitivity to SALT (Stochastic Advection by Lie Transport). The ability to generate different types of variance through either SPEC or SALT noise for TQG may have important implications in data assimilation endeavours for GFD \cite{cotter2020particle}. 
\bigskip

\textbf{Plan of the paper.}
\begin{enumerate}
    \item In \cref{sec:Equations}, we introduce the deterministic TQG equations (\cref{sec:Deterministic TQG}), and remark on their Hamiltonian formulation in \cref{sec:Hamiltonian formulation}. \Cref{Example: Ocean Bathymetry Data}  numerically demonstrates a strong correlation between bathymetry and buoyancy in the long-time dynamics. This strong correlation indicates that bathymetry is closely linked to the buoyancy variable, thereby motivating a stochastic parametrisation of bathymetry when attempting to quantify uncertainty in solution behaviour. In \cref{sec:Stochastic bathymetry} we introduce the stochastic perturbation to the bathymetry, and explain its origin via a stochastic perturbation to the potential energy in the Hamiltonian \cref{sec:Derivation of the system}. Finally, we introduce the SALT-SPEC-TQG equations in \cref{sec:SALT SPEC TQG}. 
    \item In \cref{sec:numerical study} we discuss the numerical method employed to discretise the SALT-SPEC-TQG equations. The discretisation of the deterministic spatial scheme is described in \cref{sec:numerical methods}, and the stochastic extension to SALT and SPEC is described in \cref{sec: Temporal discretisation}. In \cref{sec:deterministic} we present a deterministic solution under the initial conditions used in \cite{crisan2023btheoretical}. In \cref{sec:sensitivity study} a sensitivity to noise experiment is produced.
    \item Finally, \cref{sec: overview and outlook} presents an overview of the present work and opportunities for future research. A mathematical Appendix is also included. The Appendix illustrates the compatibility of the SALT and SPEC approaches for introducing structure-preserving stochasticity into fluid dynamics and other similar dynamical systems. This compatibility is based on the semidirect-product Lie-Poisson nature of stochastic coadjoint dynamical systems \cite{holm1998euler,marsden1984semidirect}. Examples of compatible SALT and SPEC approaches are presented for the Euler fluid equations and the stochastic dynamics of a heavy top in the Appendix, as obtained via  the Euler--Poincar\'e theory of reduction by symmetry in Lie group invariant variational principles. 
\end{enumerate}

\section{Equations}\label{sec:Equations}

\subsection{Deterministic TQG}\label{sec:Deterministic TQG}

The deterministic Thermal Quasi-Geostrophic equations as defined in \cite{crisan2023atheoretical} are given by
\begin{align}
\partial_t q + \b u \cdot \nabla (q-b) +\b u_{h}\cdot \nabla b  &= 0, \label{eq:vorticity}\\
\partial_t b + \b u \cdot \nabla b &= 0, \label{eq:bouyancy}\\
\b u = \nabla^{\perp}\psi, \quad 
\b u^{h} = \frac{1}{2}\nabla^{\perp}h_1, &\quad q = (\Delta - 1)\psi + f_1.  \label{eq:eliptic}
\end{align}
Here, $b(\b x,t)$ denotes buoyancy, $q(\b x,t)$ denotes potential vorticity, $\psi(\b x,t)$ denotes stream function.
As explained in \cite{holm2021stochastictqg,crisan2023atheoretical}, $h_1(\b x)$ is the spatial variation about a constant bathymetry, and $f_1(\b x)$ denotes the spatial variation around a constant background rotation rate. These are related to the usual Coriolis and bathymetry through the expressions $
h  = 1 + \text{Ro} h_1$, $f = 1 + \text{Ro} f_1$, where $\text{Ro}$ is the Rossby number, assumed small \cite{vallis2017atmospheric}. We shall in the rest of this work denote $f_1$ as $f$ and $h_1$ as $h$. The velocity and bathymetry-induced velocity fields $\b u(\b x,t), \b u_{h}(\b x,t)$ are divergence-free vector fields. The gradient is denoted $\nabla:=(\partial_x,\partial_y)$. The skew gradient is defined as $\nabla^{\perp}:=(-\partial_y,\partial_x)$, and the stream-functions $\psi,h$ are defined to be the negative of the usual stream function so that $u = -\nabla^{\perp}(-\psi) = \nabla^{\perp}\psi$.  The free surface elevation is defined to be $\Gamma:=\psi - b/2 $, \cite{crisan2023atheoretical}. The Jacobian of two scalar fields $a,b$ will be denoted $J(a,b):=\nabla^{\perp}a\cdot \nabla b = -a_y b_x +a_x b_y = -\nabla^{\perp}b\cdot \nabla a = J(b,a)$, and allows \cref{eq:vorticity,eq:bouyancy} to be written respectively as
\begin{align}
\partial_t q + J(\psi,q-b) + J(h/2,b) = 0,\label{eg:tqg-equations_J1}\\
\partial_t b + J(\psi,b)=0. \label{eg:tqg-equations_J2}
\end{align}

\subsection{Hamiltonian formulation}\label{sec:Hamiltonian formulation}
The TQG Hamiltonian given in \cite{holm2021stochastictqg} is
\begin{equation}
    H_{TQG}(q, b) = -\frac{1}{2}\int_{\Omega} (q-f)(\Delta - 1)^{-1} (q-f) + hb \, d^2x = -\frac{1}{2}\int_{\Omega} (q-f)\psi + hb d^2x.
\label{erg-Ham-def}
\end{equation}

Let $F(q,b)$ be a functional of state variables. The variational derivative $\frac{\delta F}{\delta q}$ is defined as 
\begin{align}
\scp{\frac{\delta F}{\delta q} }{ \delta q} 
:= \lim_{\epsilon\rightarrow 0}\frac{1}{\epsilon} \big( F[q+\epsilon \delta q,b] -F[q,b]\big),
\end{align}
The variations of the TQG Hamiltonian functional above are $( \frac{\delta H_{TQG}}{\delta q} , \frac{\delta H_{TQG}}{\delta b} ) = (-\psi,-h/2)$, and reveals a Hamiltonian structure to \cref{eg:tqg-equations_J1,eg:tqg-equations_J2}. The Poisson structure can be found, by considering the time derivative of an arbitrary functional of state variables $F(q,b)$ and using the skew-property of $J(g,h)$, to rearrange the $L^2$ integrated product below
\begin{align}
\frac{d}{dt}F  = \int_{\Omega} \begin{bmatrix}
    \delta F / \delta q \\
    \delta F / \delta b \\
\end{bmatrix}
^{T}
\begin{bmatrix}
    \partial_t q \\
    \partial_t b
\end{bmatrix}
d^2x
= - \int_{\Omega} 
\begin{bmatrix}
\delta F/ \delta q \\ \delta F/ \delta b
\end{bmatrix}
^{T}
\begin{bmatrix}
J(\psi , q-b) + J(h/2 , b ) \\
J(\psi,b)
\end{bmatrix} d^2 x,
\end{align}
into the following Lie-Poisson Hamiltonian form of the deterministic TQG equation,
\begin{align}
\frac{dF}{dt} = \{ F, H\} =  -\int 
\begin{bmatrix}
\delta F/ \delta q \\ \delta F/ \delta b
\end{bmatrix}^T
\begin{bmatrix}
J(\,q-b\,,\,\cdot\,) & J(\,b\,,\,\cdot\,)
\\
J(\,b\,,\,\cdot\,) & 0
\end{bmatrix}
\begin{bmatrix}
\delta H_{TQG}/ \delta q = \, -\psi \\ \delta H_{TQG}/ \delta b = \, -h/2
\end{bmatrix}
d^2x,
\label{eqn:TQG-LPB}
\end{align}
for the energy Hamiltonian in \eqref{erg-Ham-def}. Casimirs are defined as arbitrary functionals $C$ that vanish under the Lie-Poisson bracket $\lbrace C,D\rbrace=0$, for any functional $D$, not just the Hamiltonian \cref{eq:TQG-Hamiltonian}. Following Sophus Lie's original nomenclature, Casimirs are also called Distinguished Functionals in \cite{olver1993applications}. The conservation of the Casimirs
\begin{align}
C_{\Phi,\psi}(q,b) = \int \Phi(b) + q\Psi(b) d^2 x, \label{eq:casimirs}
\end{align}
arises through the degeneracy in the bracket \cref{eqn:TQG-LPB}, for differentiable functions $\Phi$ and $\Psi$. That is, the variational derivative components of the Casimirs comprise a null eigenvector for the Lie--Poisson matrix operator in \cref{eqn:TQG-LPB}. To prove that the bracket \cref{eqn:TQG-LPB} satisfies the defining Jacobi identity for such brackets, one can relate \cref{eqn:TQG-LPB} to the standard semidirect-product structure through the linear change of variables to $(\tilde{q},b) = (q+b,b)$ as mentioned in \cref{sec:transform to canonical form}. However, the TQG Hamiltonian in \cref{erg-Ham-def} does not collectivise in the $(q+b,b)$ variables. Consequently, the bracket \cref{eqn:TQG-LPB} is required for applications of TQG. One obtains conservation of the energy Hamiltonian \cref{erg-Ham-def} from setting $F=H_{TQG}$, and using the skew symmetry of the Poisson bracket.

\subsection{Example: Ocean Bathymetry Data}\label{Example: Ocean Bathymetry Data} We consider the following initial conditions, 
\begin{align}
    q_0 = 0,\quad
    b_0 = \sin(2 \pi y),\quad
    f = 0,\quad
    h = GEBCO. 
\end{align}
Where the initial condition for the bathymetry variation is defined through subsampling the GEBCO's 2023 Grid, a 15 arc second resolution global terrain model for elevation, providing ocean and land data \cite{GEBCO2023,jakobsson2020international}. This is done between [60,90] degrees in latitude and  [-15,15] degrees in longitude, then rescaled to [0,1], and any bathymetry data set above ocean sea level (land) has been set to 1, boundary conditions are taken to be periodic. 

Many features of the bathymetry of the Norwegian Basin 
are well beyond the asymptotic regime for which the TQG model is appropriate, and the implementation is far from being realistic (e.g. there is no forcing, no Norwegian Atlantic slope current (NwASC) and being two-dimensional, simplified boundary conditions). Nevertheless, we hope this example may illustrate some of the long-time effects of using rough topography data on the TQG equation. 

\begin{figure}[H]
\centering
\begin{subfigure}[t]{0.19\textwidth}
\centering
\includegraphics[width=.95\textwidth]{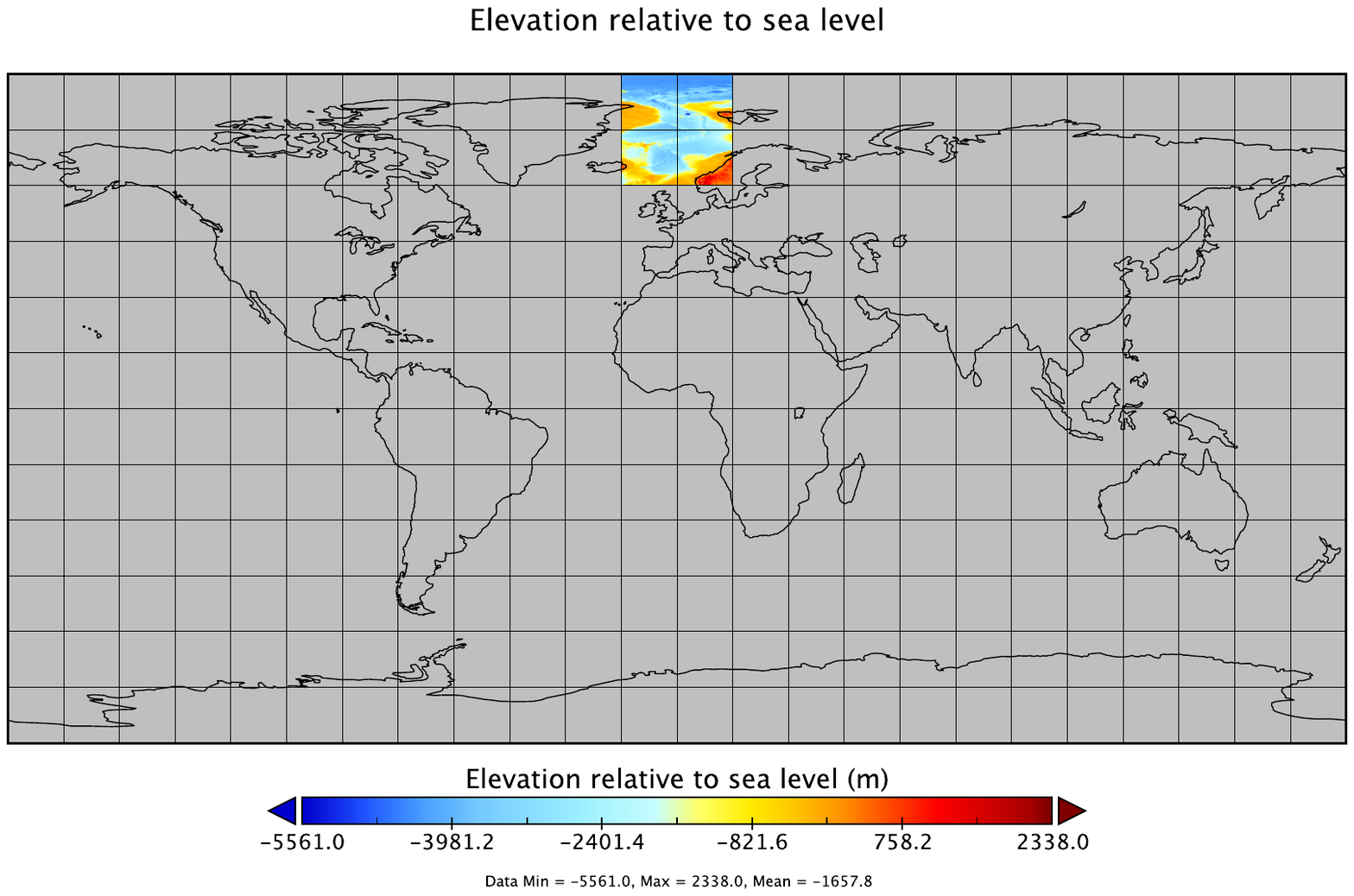}\caption{\hfill}\label{fig:bath1}
\end{subfigure}
\begin{subfigure}[t]{0.19\textwidth}
\centering
\includegraphics[width=.95\textwidth]{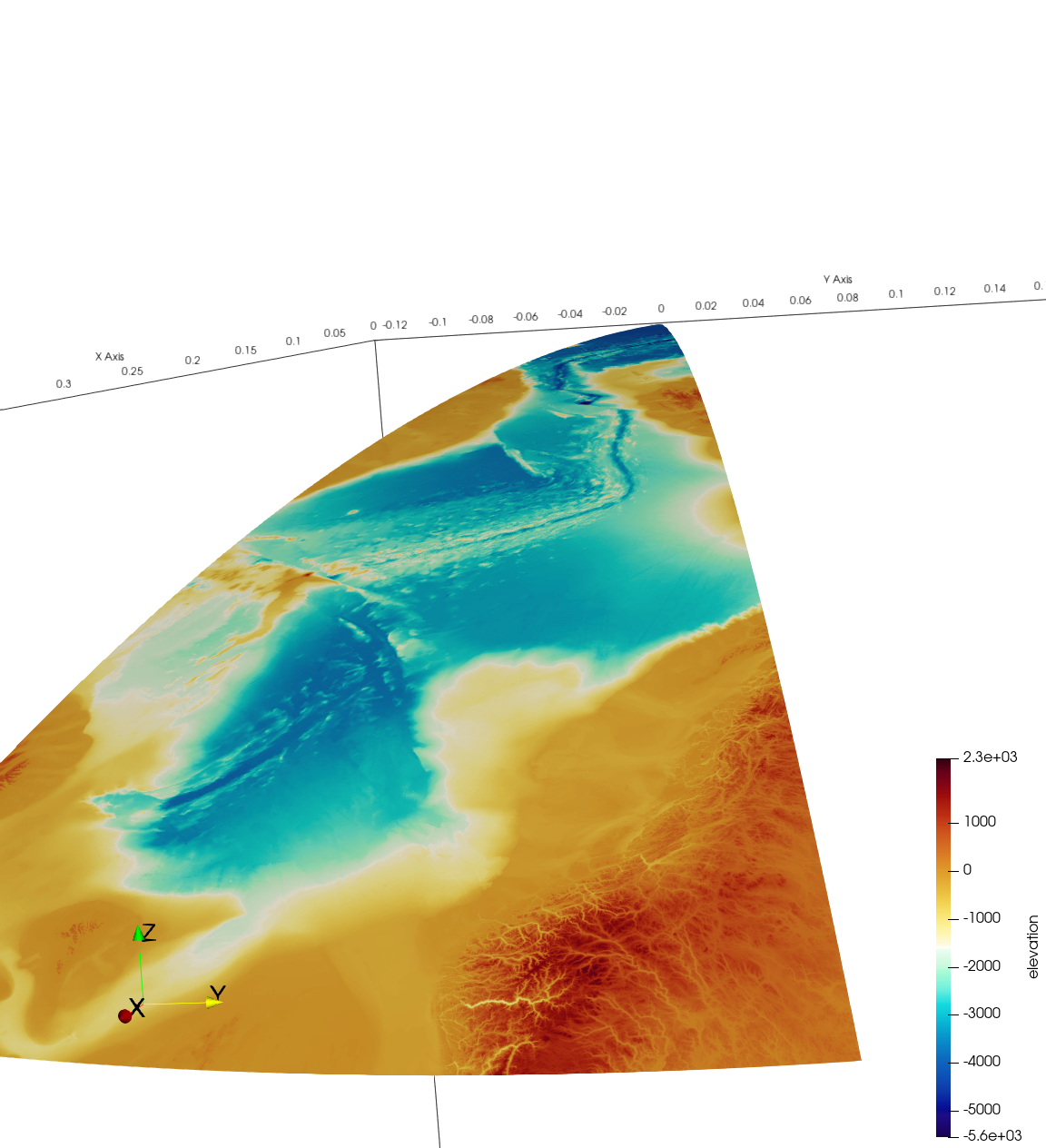}\caption{\hfill}\label{fig:bath2}
\end{subfigure}
\begin{subfigure}[t]{0.19\textwidth}
\centering
\includegraphics[width=.95\textwidth]{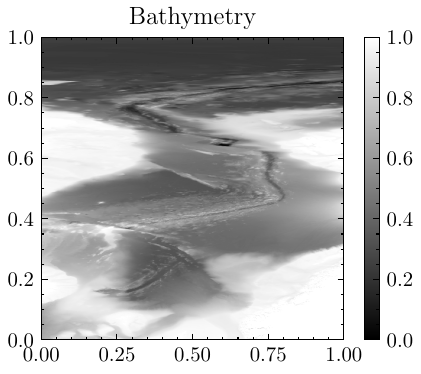}\caption{\hfill}\label{fig:bath3}
\end{subfigure}
\\
\begin{subfigure}[t]{0.19\textwidth}
\centering
\includegraphics[width=.95\textwidth]{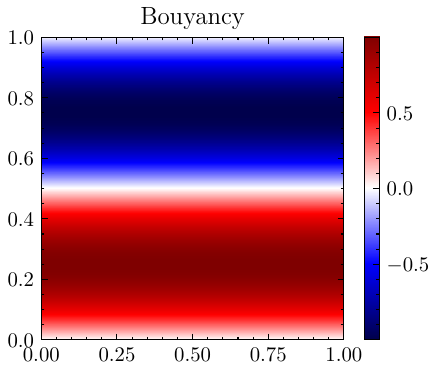}
\caption{\hfill}
\end{subfigure}
\begin{subfigure}[t]{0.195\textwidth}
\centering
\includegraphics[width=.95\textwidth]{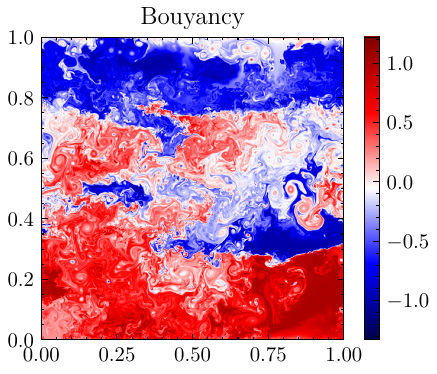}
\caption{\hfill}
\end{subfigure}
\begin{subfigure}[t]{0.195\textwidth}
\centering
\includegraphics[width=.95\textwidth]{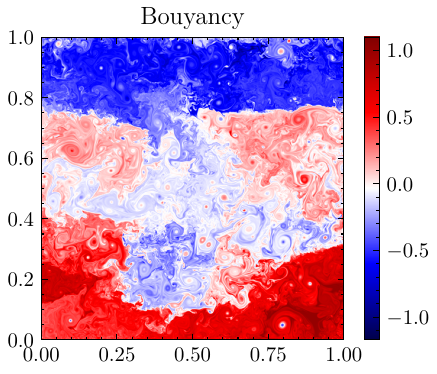}
\caption{\hfill}
\end{subfigure}
\begin{subfigure}[t]{0.195\textwidth}
\centering
\includegraphics[width=.95\textwidth]{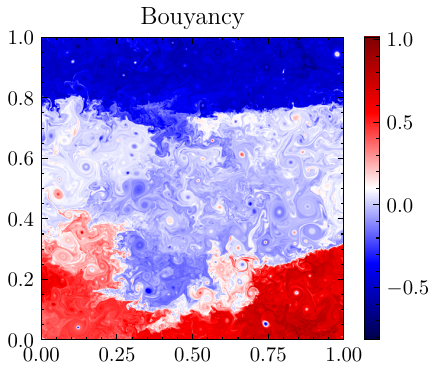}
\caption{\hfill}
\end{subfigure}
\begin{subfigure}[t]{0.195\textwidth}
\centering
\includegraphics[width=.95\textwidth]{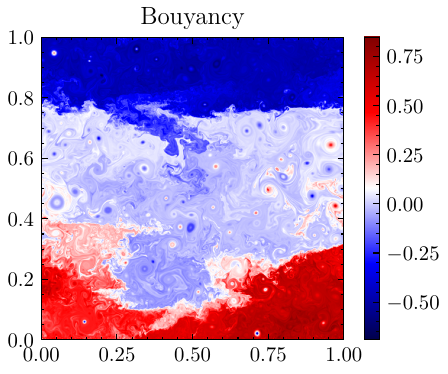}
\caption{\hfill}
\end{subfigure}
\begin{subfigure}[t]{0.19\textwidth}
\centering
\includegraphics[width=.95\textwidth]{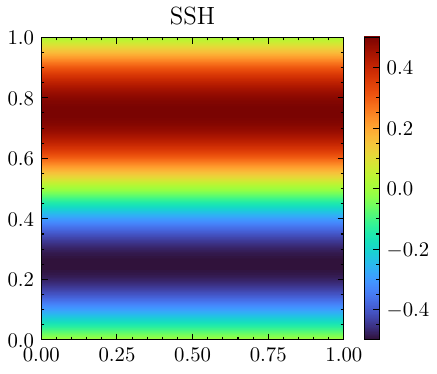}
\caption{\hfill}
\end{subfigure}
\begin{subfigure}[t]{0.195\textwidth}
\centering
\includegraphics[width=.95\textwidth]{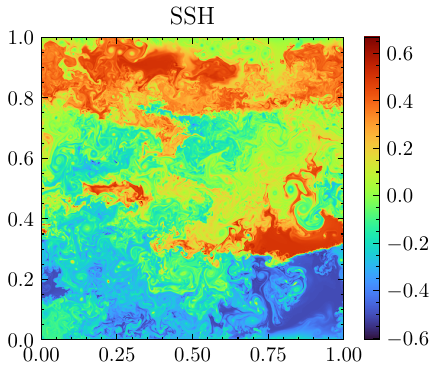}
\caption{\hfill}
\end{subfigure}
\begin{subfigure}[t]{0.195\textwidth}
\centering
\includegraphics[width=.95\textwidth]{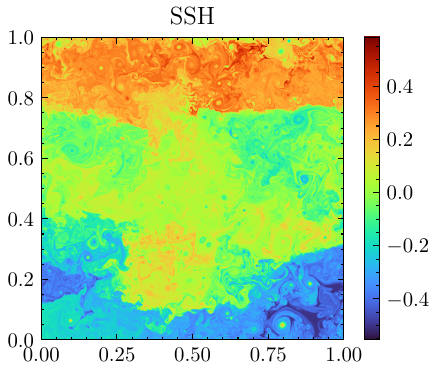}
\caption{\hfill}
\end{subfigure}
\begin{subfigure}[t]{0.195\textwidth}
\centering
\includegraphics[width=.95\textwidth]{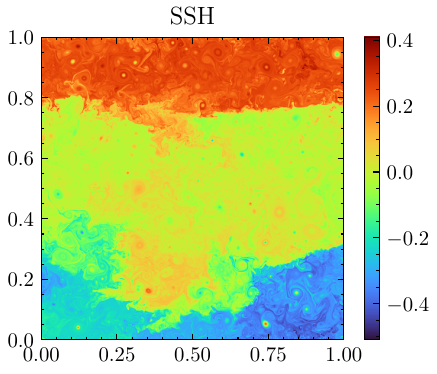}
\caption{\hfill}
\end{subfigure}
\begin{subfigure}[t]{0.195\textwidth}
\centering
\includegraphics[width=.95\textwidth]{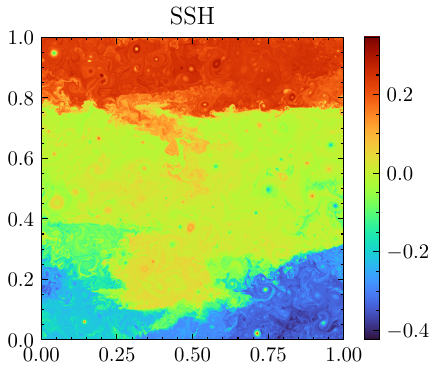}
\caption{\hfill}
\end{subfigure}
\caption{Solution of TQG over rough bathymetry. Row one, in \cref{fig:bath1}, is the bathymetry in relationship to the earth, \cref{fig:bath2} is bathymetry on the sphere, and \cref{fig:bath3} is the latitude longitude sampled bathymetry on a unit square rescaled to $[0,1]$. Row two is the Buoyancy during a run, and row three is SSH during a long time run. We observe the tendency of buoyancy and SSH to mimic the variation in bathymetry in this run.  }
\label{fig:bathymetry}
\end{figure}

We observe a tendency for buoyancy $b$ to resemble the bathymetry-variation $h$, in the long-time solution behaviour of TQG. When this occurs $b\approx h$, and the term $\nabla^{\perp}h \cdot \nabla  b \approx J(h,h) = 0$, becomes smaller. This lack of potential vorticity generation keeps the solution behaviour more stable. Potential observations include; the appearance of sharp gradients of buoyancy and SSH, and the dominant effect of bathymetry in the long time run. The features observed in \cref{fig:bathymetry} seem to indicate that the TQG buoyancy solution depends sensitively on the bathymetry.


\subsection{Stochastic bathymetry}\label{sec:Stochastic bathymetry}
We introduce the following stochastic perturbation to the TQG bathymetry, 
\begin{align}
\mathrm{d}\widetilde{h} = h(\b x) \, dt + \sum_{i}\widetilde{\zeta}_i(\b x) \circ dB_t^i, \quad \widetilde{h}(\b x, 0) = h(\b x).
\end{align}
That is, the bathymetry $h$ is replaced by  the stochastic process $\widetilde{h}$, defined by integrating the basis $\lbrace \widetilde{\zeta}_{i}\rbrace_{i=1}^{m}$ against a $m$-dimensional Wiener process $B^{i}$, $i\in \lbrace 1,...,m\rbrace$, where the notation $\circ$ indicates Stratonovich integration \cite{karatzas2014brownian}. Under this ansatz, the TQG system -- stochastically perturbed by $\mathrm{d}\widetilde{h}$ -- is given by
\begin{align}
\begin{split}
&\mathrm{d} q + \b u \cdot \nabla (q-b) \, dt + \mathrm{d}\b u_{h} \cdot \nabla b   = 0 \,,\\
&\mathrm{d} b + \b u \cdot \nabla b \, dt = 0\,,\\
&\b u = \nabla^{\perp}\psi, \quad 
\mathrm{d}\b u_{h} = \frac{1}{2}\nabla^{\perp} \mathrm{d} \widetilde{h}, \quad q - f = (\Delta - 1)\psi
\,.
\end{split}
\end{align}
These equations can be written more explicitly as follows
\begin{align}
\begin{split}
\mathrm{d} q + \b u \cdot \nabla (q-b) \, dt +\frac{1}{2}\nabla^{\perp}h \cdot \nabla b \, dt + \frac{1}{2}\sum_i\nabla^{\perp} \widetilde{\zeta}_i(\b x) \cdot \nabla b \circ dB^i_t &= 0 \,,\\
\mathrm{d} b + \b u \cdot \nabla b \, dt &= 0 \,.
\end{split}
\label{PerturbedTQG}
\end{align}

\subsection{Derivation of the system in \eqref{PerturbedTQG}}\label{sec:Derivation of the system}
We perturb the TQG Hamiltonian \cref{erg-Ham-def} by an additional time-varying noise term designed to capture uncertainty in the potential energy,
\begin{equation}
    \
    \mathrm{d} \widetilde{H}(q, b) = H_{TQG}\, dt 
    - \frac12 \sum_i \int_\D \widetilde{\zeta}_i(\bx) \, b(\bx,t) \, d^2 x \circ dW_t^i\,.
    \label{eq:TQG-Hamiltonian}
\end{equation}

The Hamiltonian Poisson bracket is derived, by considering the stochastic time derivative of an arbitrary functional of state variables, and substituting in the stochastically perturbed equations to give, 
\begin{align}
\mathrm{d}F =  -\int 
\begin{bmatrix}
\delta F/ \delta q \\ \delta F/ \delta b
\end{bmatrix}^T
\begin{bmatrix}
J(\,q-b\,,\,\cdot\,) & J(\,b\,,\,\cdot\,)
\\
J(\,b\,,\,\cdot\,) & 0
\end{bmatrix}
\begin{bmatrix}
\delta \widetilde{H} /  \delta q = - \psi \, dt\\ 
\delta \widetilde{H}/ \delta b = -\frac h2 \, dt - \frac12\sum_i \widetilde{\zeta}_i \circ dB_t^i 
\end{bmatrix}
d^2 x
\label{eqn:TQG-Euler-LPB}.
\end{align}

The stochastic Hamiltonian system in \eqref{eqn:TQG-Euler-LPB}, has the same Lie Poisson bracket as the deterministic system \eqref{eqn:TQG-Euler-LPB}, and therefore possesses the same degeneracy. Consequently, the stochastic system \eqref{eqn:TQG-Euler-LPB} conserves the same Casimirs as in the deterministic case,  \cref{eq:casimirs}.

\subsection{SALT SPEC TQG}\label{sec:SALT SPEC TQG}
As remarked in the introduction, both SALT (\cite{holm2015}) and SPEC allow for stochastic parametrisation of the TQG equations, allowing for the Casimirs to be preserved. For the SALT TQG system, we refer to \cite{crisan2023btheoretical,holm2015} for definition, motivation and Lie-Poisson structure. The following perturbation 
\begin{align}
\mathrm{d}  \widehat{\b u} = \b u dt + \sum_{i} \b\xi_{i}(\b x) \circ dW^{i}, \quad \widehat{u}(\b x,0)=\b u (\b x),
\end{align}
is considered at the level of the Lie algebra. In the Lie Poisson structure SALT noise is added to the $\delta \widetilde{H}/\delta q$ term in \cref{eqn:TQG-Euler-LPB}. Where it can be assumed that for an incompressible basis $\operatorname{div} \b \xi_{i}(\b x)=0$, $\forall i$ that $\b\xi_i = \nabla^{\perp} \Psi_i$ arises from a stream function. 
In the variables $(q,b)$ the SALT-SPEC-TQG equations are given by
\begin{align}
\rd q + (\b u dt +\sum_{i} \b \xi_{i}(x)\circ dW^{i}) \cdot \nabla (q-b)  + (\b u_h dt + \sum_{i}\b \eta_{i} \circ dB^{i}) \cdot \nabla b  &= 0, \label{eq:salt-spec-tqg1}\\
\rd b + (\b u dt+\sum_{i}\b \xi_{i}(x)\circ dW^{i})\cdot \nabla b &= 0. \label{eq:salt-spec-tqg2}
\end{align}
Where $\b \eta_i = 1/2\nabla^{\perp}\tilde{\zeta}_i$, $q - f = (\Delta - 1)\psi$, $\b u=\nabla^{\perp}\psi$, $\b\xi_i = \nabla^{\perp} \Psi_i$, $\b u_{h} = \nabla^{\perp} h_0/2$. For the compatibility of SALT and SPEC approaches see \cref{sec:Introducing two types of stochasticity}. We note that SALT perturbations directly transport both $q,b$, whereas SPEC peturbations only directly effect $q$, through the ``forcing" multiplying the gradient of bathymetry, and effect $b$ only through the coupling in the system. 

\section{Numerical Study}\label{sec:numerical study}
\subsection{Numerical Methods}\label{sec:numerical methods}
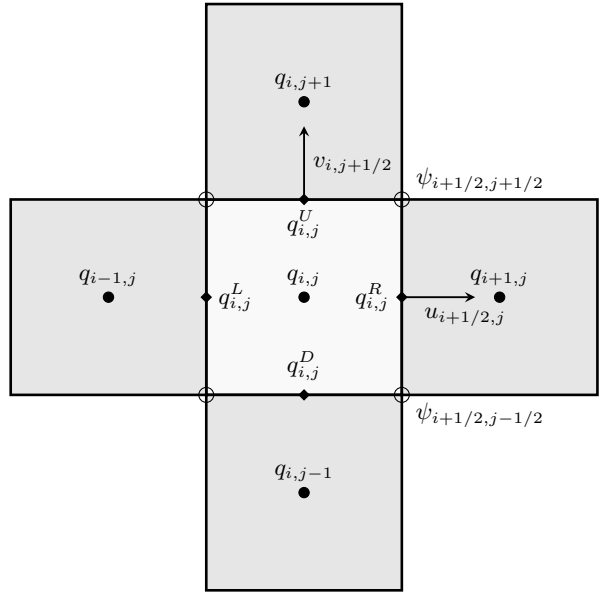
\begin{figure}[H]
\centering
\begin{subfigure}[b]{0.45\textwidth}
\begin{tikzpicture}[scale=1.3]
\draw[fill=gray!5,line width=1] (5,3) rectangle (7,5);
\draw[fill=gray!20,line width=1] (5,5) rectangle (7,7);
\draw[fill=gray!20,line width=1] (5,1) rectangle (7,3);
\draw[fill=gray!20,line width=1] (3,3) rectangle (5,5);
\draw[fill=gray!20,line width=1] (7,3) rectangle (9,5);
\draw[-stealth,line width=0.25mm] (7,4)--(7.75,4);
\draw[-stealth,line width=0.25mm] (6,5)--(6,5.75);
\node at (6.7,4)  {\small{$q^{R}_{i,j}$}};
\node at (6,4.2) {\small{$q_{i,j}$}};
\node at (5.3,4) {\small{$q^{L}_{i,j}$}};
\node at (6,4.75) {\small{$q^{U}_{i,j}$}};
\node at (6,3.3) {\small{$q^{D}_{i,j}$}};
\filldraw (6,4) circle[radius=1.5pt];
\filldraw (8,4) circle[radius=1.5pt];
\filldraw (4,4) circle[radius=1.5pt];
\filldraw (6,6) circle[radius=1.5pt];
\filldraw (6,2) circle[radius=1.5pt];
\node at (8,4.2) {\small{$q_{i+1,j}$}};
\node at (4,4.2) {\small{$q_{i-1,j}$}};
\node at (6,6.2) {\small{$q_{i,j+1}$}};
\node at (6,2.2) {\small{$q_{i,j-1}$}};
\node [draw, circle, scale=0.15] at (7,5) {$corner$};
\node at (7.8,5.2) {\small{$\psi_{i+1/2,j+1/2}$}};
\node [draw, circle, scale=0.15] at (7,3) {$corner$};
\node at (7.8,2.8) {\small{$\psi_{i+1/2,j-1/2}$}};
\node [draw, circle, scale=0.15] at (5,5) {$corner$};
\node [draw, circle, scale=0.15] at (5,3) {$corner$};

\node [draw, diamond, fill, scale=0.125] at (6,3) {$ok$};
\node [draw, diamond, fill, scale=0.125] at (6,5) {$ok$};
\node [draw, diamond, fill, scale=0.125] at (7,4) {$ok$};
\node [draw, diamond, fill, scale=0.125] at (5,4) {$ok$};
\node at (6.5,5.35) {\small{$v_{i,j+1/2}$}};
\node at (7.65,3.8) {\small{$u_{i+1/2,j}$}};
\end{tikzpicture}
\caption{Flux form stencil.}
\label{fig:flux form stencil}
\end{subfigure}
\caption{Diagram of several variables on the C-grid, and subcell reconstruction values. $q_{i,j},b_{i,j}$ belong at cell center. $q,b,(q-b)$ are reconstructed at Up, Down, Left and Right of the cell denoted with the superscripts $U,D,L,R$, respectively allowing for edge discontinuity. Streamfunctions $\psi, h$ are located at the corners and used in the computation of the divergence free velocity fields $(u_{i+1/2,j},v_{i,j+1/2})$, $(u^{h}_{i+1/2,j},v^{h}_{i,j+1/2})$. }
\end{figure}

 Using the divergence-free property it is possible to write \cref{eq:bouyancy,eq:vorticity} in flux form as follows
\begin{align}
    \frac{d q}{dt} + \nabla \cdot (\bu (q-b))+\nabla \cdot (\bu_{h} q ) & = 0, \label{eq:flux_form_vorticity}\\
    \frac{d b}{dt} + \nabla \cdot (\bu b) & = 0. \label{eq:flux_form_bouyancy}
\end{align}
Such that a linear invariant, locally conservative, consistent, flux form scheme can be employed \cite{woodfield2024new}. Using a C-grid style implementation (\cref{fig:flux form stencil})\cite{arakawa1981potential}, it is possible to construct a discrete skew gradient differential operator $\nabla^{\perp}_{h}$ through an interpolation $\mathcal{I}_h$ to corner points, and a discrete divergence operator $\text{div}_{h}$, capable of ensuring $\text{div}_{h}\nabla^{\perp}_{h}f = 0 $ to machine precision. This is a mimetic flux form discretisation, capable of locally conserving flux of potential vorticity (and buoyancy). The elliptic relation is solved spectrally. Upwinding allows built-in diffusion. Weights can be chosen such that for unidirectional flow, arbitrary order convergence is attained whilst retaining the one-point second-order cheap Gauss quadrature in the evaluation of the flux. This is described in the semi-discrete deterministic context below. 

\begin{method}[Semi-Discrete method]
\phantom{ok}
\begin{enumerate}
\item Solve for the streamfunction $\psi$ in spectral space from $q,f$, this is done using the 2D fast Fourier transform \cite{cooley1965algorithm} and its inverse (denoted by $\mathcal{F}$ and $\mathcal{F}^{-1}$ respectively) as follows 
\begin{align}
\psi &=\mathcal{F}^{-1} \left( \frac{\mathcal{F}(q-f)}{[(2\pi i k_x)^2+(2\pi i k_y)^2-1]} \right).  \label{eq:spectral}
\end{align}
Where $k_x,k_y$ are the wavenumbers in the $x$ and $y$ direction respectively. 
\item Interpolate stream function and bathymetry to the corner points $(i,j)\mapsto(i+1/2,j+1/2)$, as follows 
\begin{align}
\psi_{i+1/2,j+1/2} &= (\psi_{i,j}+ \psi_{i+1,j}+\psi_{i,j+1} + \psi_{i+1,j+1})/4, \label{eq:interpolation} \\
h_{i+1/2,j+1/2} &= (h_{i,j}+ h_{i+1,j}+h_{i,j+1} + h_{i+1,j+1})/4.
\end{align}
\item Solve for the velocity fields $(u,v)= \nabla^{\perp} \psi$, $(u_{h},v_{h})= \frac{1}{2}\nabla^{\perp} h$ with second order differences 
    \begin{align}
        u_{i+1/2,j} &= -\frac{\psi_{i+1/2,j+1/2} - \psi_{i+1/2,j-1/2}}{\Delta y},\quad 
        v_{i,j+1/2} = \frac{\psi_{i+1/2,j+1/2} - \psi_{i-1/2,j+1/2}}{\Delta x},\label{eq: corner grad}\\
        u^{h}_{i+1/2,j} &= -\frac{h_{i+1/2,j+1/2} - h_{i+1/2,j-1/2}}{2\Delta y}, \quad
        v^{h}_{i,j+1/2} = \frac{h_{i+1/2,j+1/2} - h_{i-1/2,j+1/2}}{2\Delta x}.
    \end{align}
    \item We discretise \cref{eq:flux_form_vorticity} in flux form as follows
    \begin{align}
    \frac{d q}{dt} &=  -\frac{1}{\Delta x}(F^{q-b}_{i+1/2,j} - F^{q-b}_{i-1/2,j}) - \frac{1}{\Delta y}(F^{q-b}_{i,j+1/2} - F^{q-b}_{i,j-1/2}),\\
    &\phantom{=} - \frac{1}{\Delta x}(F^{h}_{i+1/2,j} - F^{h}_{i-1/2,j}) - \frac{1}{\Delta y}(F^{h}_{i,j+1/2} - F^{h}_{i,j-1/2}),\\
    \frac{d b}{dt} &=  -\frac{1}{\Delta x}(F^b_{i+1/2,j} - F^b_{i-1/2,j}) - \frac{1}{\Delta y}(F^b_{i,j+1/2} - F^b_{i,j-1/2}),
    \end{align}
    where the flux at the right edge of each cell is resolved using the donor cell numerical flux as follows
    \begin{align}
    F^b_{i+1/2,j} &= u_{i+1/2,j}^+ b^{R}_{i,j} + u_{i+1/2,j}^- b^{L}_{i+1,j},\\
    F^{q-b}_{i+1/2,j} &= u_{i+1/2,j}^{+} (q -b )^{R}_{i,j} + u_{i+1/2,j}^{-} (q -b)^{L}_{i+1,j},\\
    F^{h}_{i+1/2,j} &= (u^{h}_{i+1/2,j})^+ q^{R}_{i,j} + (u^{h}_{i+1/2,j})^- q^{L}_{i+1,j}, \\
    F^{b}_{i+1/2,j} &= u_{i+1/2,j}^{+} b^{R}_{i,j} + u_{i+1/2,j}^{-} b^{L}_{i+1,j},
    \end{align}
    where
    $u_{i+1/2,j}^+$, $u_{i+1/2,j}^-$ denote the positive and negative $x$-component (normal component on a Cartesian grid) of velocity vector at face $(i+1/2,j)$ respectively, where $q^{R}_{i,j}$ denotes the subcell value of $q_{i,j}$ at position $(x_{i+1/2},y_{j})$. $q^{L}_{i+1,j}$ denotes the subcell value of $q_{i+1,j}$, at position $(x_{i+1/2},y_{j})$.  
\item The interpolated values are constructed such that the flux difference is considered higher order in the pointwise finite difference sense for uniform flow, and second order in a finite volume sense approximating the integral form of the equation. For example the reconstruction, 
\begin{align}
    q^{R}_{i,j} = w_{k}q_{i+k} + w_{k-1}q_{i+k-1} ... + w_{-k}q_{i-k}, \label{eq:interpolation weights}
\end{align}
for uniform positive flow leads to the flux difference
\begin{align}
\frac{1}{\Delta x} [q^{R}_{i,j} - q^{R}_{i-1,j}]  = \frac{1}{\Delta x} [(w_{k})q_{i+k} + (w_{k}-w_{k-1})q_{i+k-1} ... - w_{-k-1}q_{i-k-1}] = O(\Delta x^{2k+1}). \label{eq:finite difference weights}
\end{align}
Upon appropriate choice of weights. The computation of the finite difference weights (in \cref{eq:finite difference weights}) is done before runtime using \cite{fornberg1998classroom}, and turned into flux interpolation weights $w_{k}$ in \cref{eq:interpolation weights}. This generalises to Cartesian meshes \cite{fornberg1998classroom} and can be implemented with linear computational cost using the second form of barycentric polynomial interpolation formula \cite{trefethen2019approximation}.
\end{enumerate}
\end{method}
The flux building weights $\lbrace w_{k}\rbrace_{k=-4}^{4}=
1/2520[-5, 55, -305, 1375, 1879, -641,  199,  -41,    4],$
attains the order 9 upwind bias stencil,
$
1/2520[-5,60,-360,1680,504,-2520,840,-240,45,-4]$. In practice the flux building weights are used $\lbrace w_k \rbrace_{k=-2}^{2} = 1/60 [-3,27,47,-13,2]$ resulting in the 5th order upwind stencil,
$
1/60[-3,30,20,-60,15,-2]$.

\subsection{Temporal discretisation}\label{sec: Temporal discretisation}
We discretise the stochastic (SALT-SPEC) equations using a stochastic generalisation of the Strong Stability Preseving (SSP) three-stage third-order Runge-Kutta scheme SSP33. Originally appearing in its deterministic form in \cite{shu1988efficient}, the stochastic generalisation is found by replacing the Forward Euler stages in the Shu-Osher representation by the Euler Maruyama (EM) scheme. Note that in this setting, the scheme is not necessarily strong stability preserving, and is implemented due to its low memory considerations.

\begin{algorithm}[SSP33-SALT-SPEC-TQG]
\begin{algorithmic}
Assign
\begin{align}
(q^{(1)},b^{(1)}) & =\text {EM}(q^{n}, b^{n},\Delta t, \Delta B^k,\Delta W^k,...), \\
(q^{(2)},b^{(2)}) & =3 / 4 (q^{n}, b^{n})  +1 / 4 \text{EM}(q^{(1)}, b^{(1)},\Delta t, \Delta B^k,\Delta W^k,...), \\
(q^{n+1},b^{n+1}) & =1 / 3 (q^{n}, b^{n}) + 2 / 3 \text{EM}(q^{(2)}, b^{(2)},\Delta t, \Delta B^k,\Delta W^k,...).
\end{align}

\RETURN $(q^{n+1},b^{n+1})$
\end{algorithmic}
\end{algorithm}
The SALT and SPEC stochastic perturbations are combined in the EM scheme assuming stream-function representations as follows.

\begin{algorithm}[EM-SALT-SPEC-TQG]
\label{alg:buildtree}
\begin{algorithmic}
\STATE{We describe $EM(q^{(s)}, b^{(s)},\Delta t, \Delta B^k,\Delta W^k,...) $}
\STATE{Set $\psi^{(s)}_{i,j} =\mathcal{F}^{-1} \left( \frac{\mathcal{F}(q^{(s)}-f)}{[(2\pi i k_x)^2+(2\pi i k_y)^2-1]} \right),\quad \forall i,j$, as in \cref{eq:spectral}.  }
\STATE{Set $
\psi^{(s)}_{1,i,j}= \psi_{i,j}^{(s)} + \sum_{k=1}^{m} \Psi_{k}(\b x_{i,j}) \Delta W^{k} / \Delta t$, }
\STATE{Set 
$\psi^{(s)}_{2,i,j} = 1/2 h_0(\b x_{i,j}) + 1/2 \sum_{k=1}^{m}\widetilde{\zeta}_{k}(\b x_{i,j}) \Delta B^{k}/\Delta t.$}
\STATE{Compute $\psi^{(s)}_{l,i+1/2,j+1/2} = I_{h} \circ \psi^{(s)}_{l,i,j}$, for $l=\lbrace 1,2\rbrace$ using \cref{eq:interpolation}}
\STATE{Compute $\b u_{l}^{(s)},= \nabla^{\perp}_{h}\psi_{l}^{(s)}$ for $l=\lbrace 1,2\rbrace$ using \cref{eq: corner grad}.}
\STATE{Update
$(q^{(s+1)},b^{(s+1)}) = (q^{(s)} + \Delta t \operatorname{div}_{h}( (q^{(s)}-b^{(s)})\b u^{(s)}_2) + \Delta t \operatorname{div}_{h}(q^{(s)}\b u^{(s)}_1), b^{(s)} +  \Delta t \operatorname{div}_{h}(b^{(s)}\b u^{(s)}_1)) $} using a forward Euler discretisation of the semidiscrete scheme.
\RETURN $(q^{(s+1)},b^{(s+1)})$
\end{algorithmic}
\end{algorithm}

Since the deterministic scheme SSP33, captures second-order (and higher order) terms, the stochastic generalisation captures the Itô-Stratonovich correction in the Stratonovich-Taylor expansion, and the scheme can be shown to be weak order 1, or strong order 1/2 by Taylor expanding. The temporal consistency arguments are a subcase of those presented in \cite{ruemelin1982numerical}. Details of temporal consistency in reference to the SALT-TQG system can be found in \cite{crisan2023btheoretical}.



\subsection{Deterministic}\label{sec:deterministic}

\begin{figure}[H]
\centering
\begin{subfigure}[t]{0.245\textwidth}
\centering
\includegraphics[width=.95\textwidth]{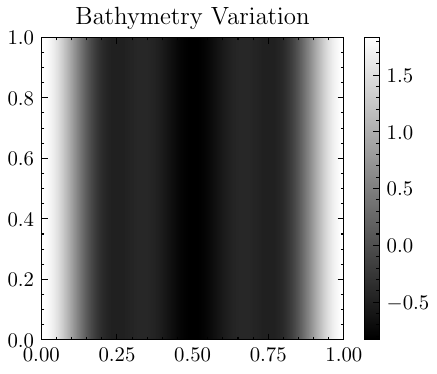}
\caption{\hfill}\label{fig:Bathymetry}
\end{subfigure}
\begin{subfigure}[t]{0.245\textwidth}
\centering
\includegraphics[width=.95\textwidth]{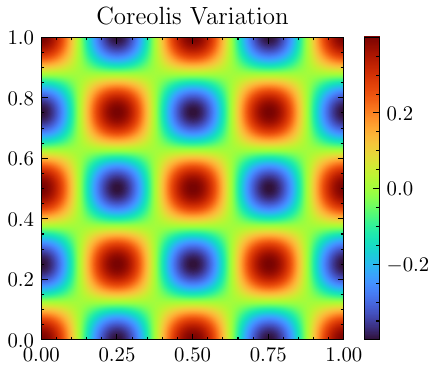}
\caption{\hfill}\label{fig:Coreolis}
\end{subfigure}
\begin{subfigure}[t]{0.245\textwidth}
\centering
\includegraphics[width=.95\textwidth]{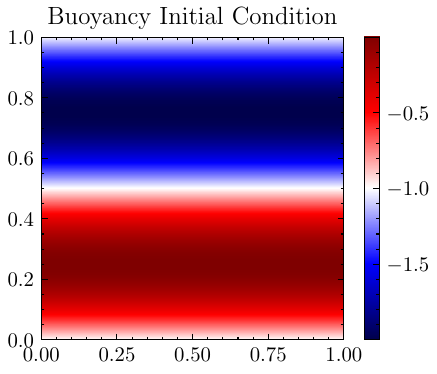}
\caption{\hfill}\label{fig:BuoyancyIC}
\end{subfigure}
\begin{subfigure}[t]{0.245\textwidth}
\centering
\includegraphics[width=.95\textwidth]{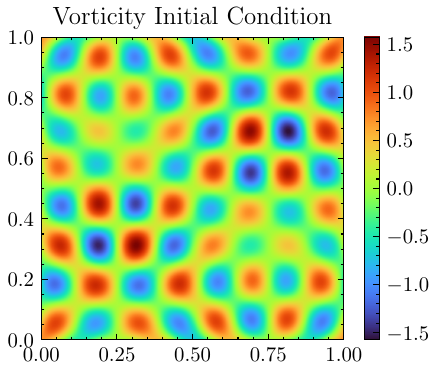}
\caption{\hfill}\label{fig:VorticityIC}
\end{subfigure}
\caption{Initial conditions $(h,f,b_0,q_0)$ from \cite{crisan2023atheoretical}.}
\label{fig:initial conditions crisan}
\end{figure}

We consider the initial conditions proposed in \cite{crisan2023btheoretical}, defined as follows 
\begin{align}
q(0, x, y)&=  \sin (8 \pi x) \sin (8 \pi y)+0.4 \cos (6 \pi x) \cos (6 \pi y) +0.3 \cos (10 \pi x) \cos (4 \pi y) +0.02 \sin (2 \pi y)+0.02 \sin (2 \pi x), \label{eq:ic:vorticity}\\
b(0, x, y)&=  \sin (2 \pi y)-1,\label{eq:ic:Bouyancy}\\
 h(x, y)&=\cos (2 \pi x)+\frac{1}{2} \cos (4 \pi x)+\frac{1}{3} \cos (6 \pi x), \label{eq:ic:bathymetry}\\
f(x, y) &= 0.4 \cos (4 \pi x) \cos (4 \pi y),\label{eq:ic:coreolis}
\end{align}
on the unit tourus $\mathbb{T}^2$, plotted in \cref{fig:initial conditions crisan}. 

\begin{figure}[!hbt]
\centering
\begin{subfigure}[t]{0.245\textwidth}
\centering
\includegraphics[width=.95\textwidth]{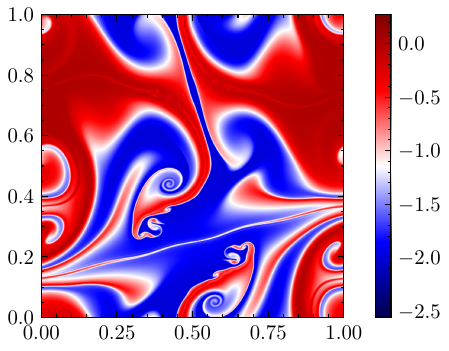}
\caption{\hfill}\label{fig:highres1}
\end{subfigure}
\begin{subfigure}[t]{0.245\textwidth}
\centering
\includegraphics[width=.95\textwidth]{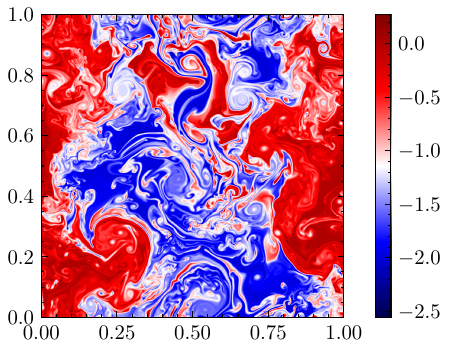}
\caption{\hfill}\label{fig:highres2}
\end{subfigure}
\begin{subfigure}[t]{0.245\textwidth}
\centering
\includegraphics[width=.95\textwidth]{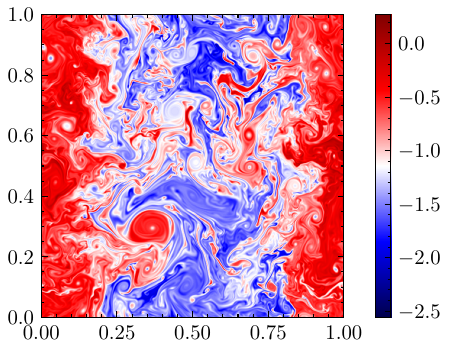}
\caption{\hfill}
\end{subfigure}
\begin{subfigure}[t]{0.245\textwidth}
\centering
\includegraphics[width=.95\textwidth]{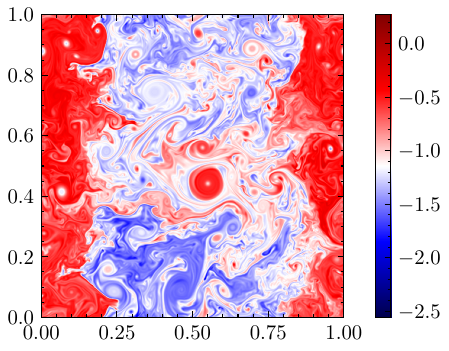}
\caption{\hfill}
\end{subfigure}\\
\begin{subfigure}[t]{0.245\textwidth}
\centering
\includegraphics[width=.95\textwidth]{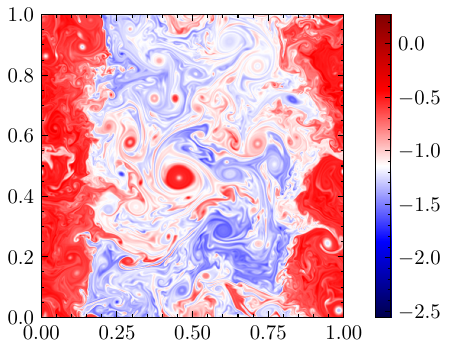}
\caption{\hfill}
\end{subfigure}
\begin{subfigure}[t]{0.245\textwidth}
\centering
\includegraphics[width=.95\textwidth]{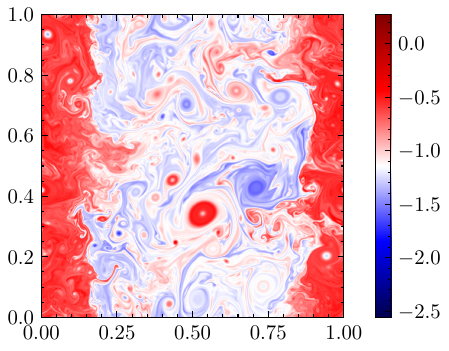}
\caption{\hfill}
\end{subfigure}
\begin{subfigure}[t]{0.245\textwidth}
\centering
\includegraphics[width=.95\textwidth]{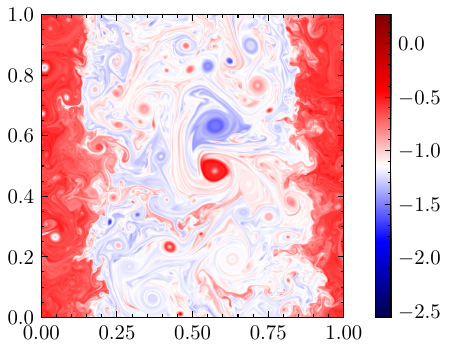}
\caption{\hfill}
\end{subfigure}
\begin{subfigure}[t]{0.245\textwidth}
\centering
\includegraphics[width=.95\textwidth]{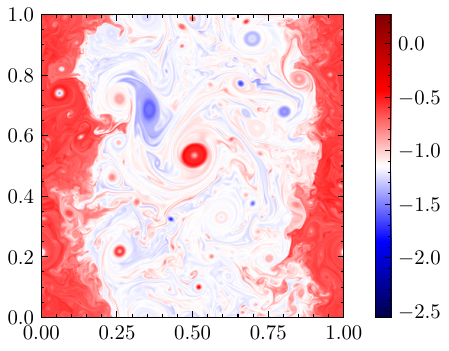}
\caption{\hfill}
\end{subfigure}
\caption{Buoyancy $b$ at times $t\in 1.25[8]$ at $512\times512$ spatial resolution, in consequtive order. Initial conditions from \cite{crisan2023atheoretical}.}
\label{fig:highres buoyancy}
\end{figure}

In the high-resolution experiment we use grid resolution of $n_x = n_y = 512$, and run with $n_t = 20480$ timesteps over the interval $t\in[0,10]$, resulting in a timestep of $\Delta t = 0.00048828125$. 
We will denote the set of natural numbers from 1 to n, with the square bracket $[n]:=\lbrace 1,...n \rbrace$,  such that the discrete-time window on $t\in[0,T]$ used for time-stepping is denoted $\mathcal{T}_{n_t,\Delta t}:= \lbrace \Delta t, 2\Delta t ...,T \rbrace = \Delta t [n_t]$. The buoyancy at $t\in 1.25[8]$ is plotted in \cref{fig:highres buoyancy}. We observe spin up behaviour in the first two plots in \cref{fig:highres1,fig:highres2}, and subsequently in the remaining six figures in \cref{fig:highres buoyancy} the TQG buoyancy resembles bathymetry variation \cref{fig:Bathymetry}. 

\subsection{Sensitivity study}\label{sec:sensitivity study}
In the lower resolution SALT and SPEC ensembles we use grid resolution $n_x = n_y = 128$, with $n_t = 5120$ on $t\in[0,10]$ resulting in $\Delta t = 0.001953125$. We use $E=64$ ensemble members in both ensembles, each ensemble member uses the same $P=64=8\times 8$ basis functions,
\begin{align}
\Psi_{r,s} = \sin(2 r \pi x)\sin(2 s \pi y)/(rs), \quad \forall (r,s) \in [8]\times [8], \\
\widehat{\zeta}_{r,s} = \sin(2 r \pi x)\sin(2 s \pi y)/(rs), \quad \forall (r,s) \in [8]\times [8].
\end{align}
 
\begin{figure}[H]
\centering
\begin{subfigure}[t]{0.245\textwidth}
\centering
\includegraphics[width=.95\textwidth]{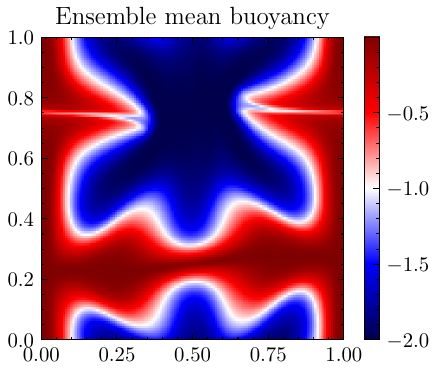}
\caption{\hfill}
\end{subfigure}
\begin{subfigure}[t]{0.245\textwidth}
\centering
\includegraphics[width=.95\textwidth]{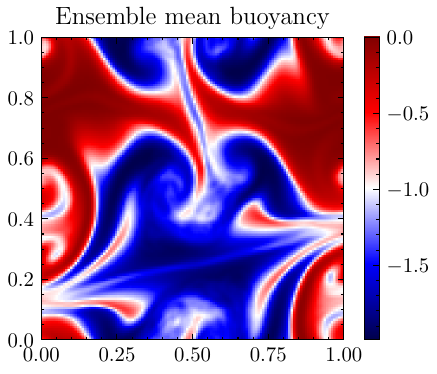}
\caption{\hfill}
\end{subfigure}
\begin{subfigure}[t]{0.245\textwidth}
\centering
\includegraphics[width=.95\textwidth]{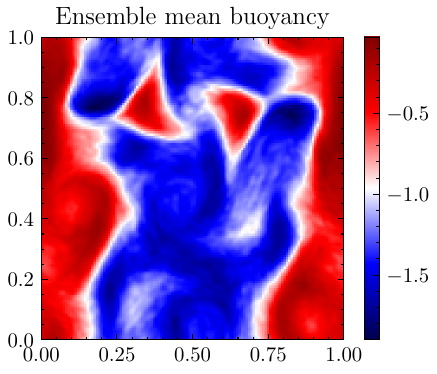}
\caption{\hfill}
\end{subfigure}
\begin{subfigure}[t]{0.245\textwidth}
\centering
\includegraphics[width=.95\textwidth]{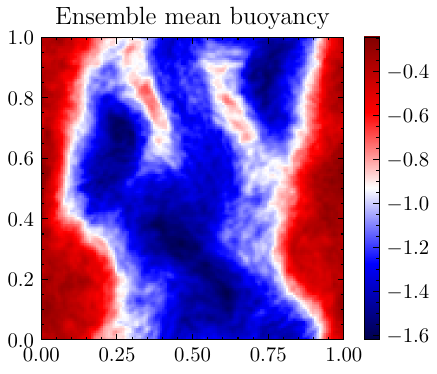}
\caption{\hfill}
\end{subfigure}
\\
\begin{subfigure}[t]{0.245\textwidth}
\centering
\includegraphics[width=.95\textwidth]{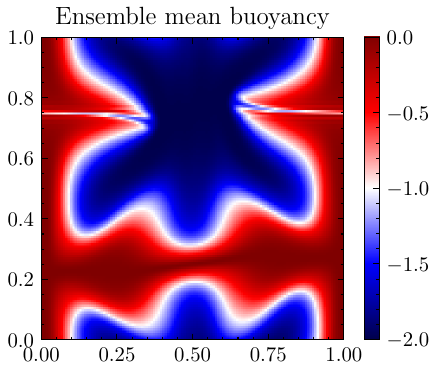}
\caption{\hfill}
\end{subfigure}
\begin{subfigure}[t]{0.245\textwidth}
\centering
\includegraphics[width=.95\textwidth]{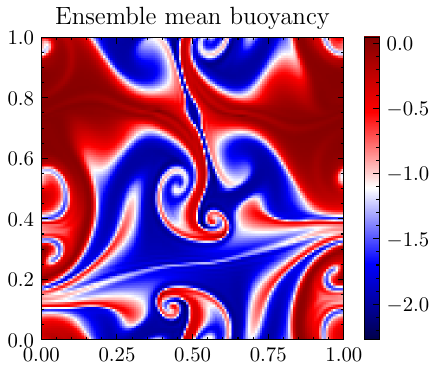}
\caption{\hfill}
\end{subfigure}
\begin{subfigure}[t]{0.245\textwidth}
\centering
\includegraphics[width=.95\textwidth]{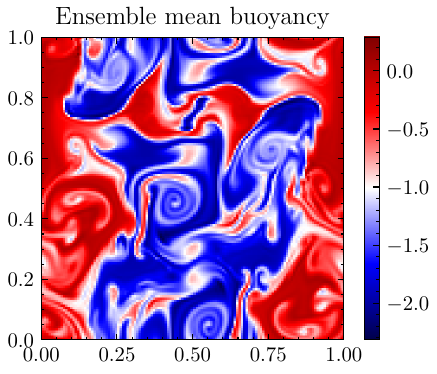}
\caption{\hfill}
\end{subfigure}
\begin{subfigure}[t]{0.245\textwidth}
\centering
\includegraphics[width=.95\textwidth]{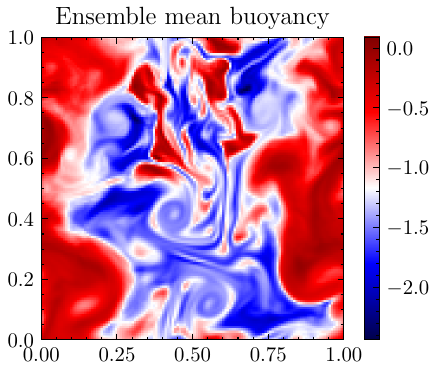}
\caption{\hfill}
\end{subfigure}
\caption{Row 1, the ensemble mean of a 64 member SALT ensemble at times $t \in 0.625 [4]:=\lbrace 0.625,1.25,1.875,2.5\rbrace$, during spin-up. Row 2, the ensemble mean of a 64-member SPEC ensemble at times $t \in 0.625 [4]$, during spin-up. Finer structures can be observed in the SPEC ensemble, indicating ensemble members share similar features.}
\label{fig:EnsembleMean}
\end{figure}

\begin{figure}[!hbt]
\centering
\begin{subfigure}[t]{0.245\textwidth}
\centering
\includegraphics[width=.95\textwidth]{Figs_JW/SALTEnsembleMeanBuoyancy_1280.pdf}
\caption{\hfill}
\end{subfigure}
\begin{subfigure}[t]{0.245\textwidth}
\centering
\includegraphics[width=.95\textwidth]{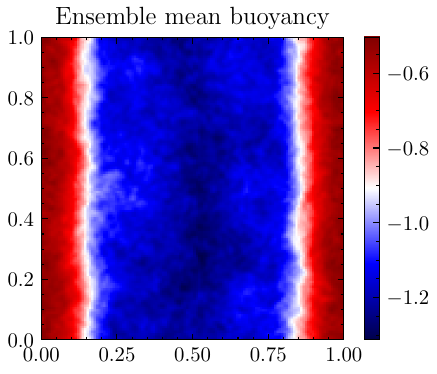}
\caption{\hfill}
\end{subfigure}
\begin{subfigure}[t]{0.245\textwidth}
\centering
\includegraphics[width=.95\textwidth]{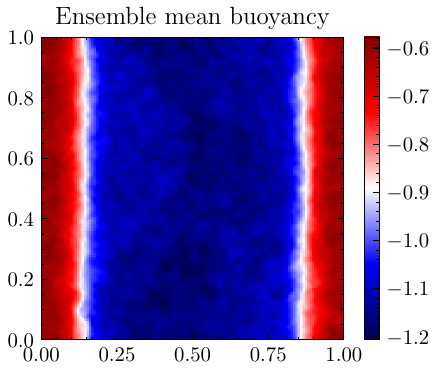}
\caption{\hfill}
\end{subfigure}
\begin{subfigure}[t]{0.245\textwidth}
\centering
\includegraphics[width=.95\textwidth]{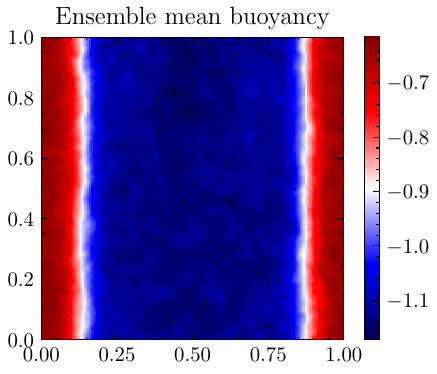}
\caption{\hfill}
\end{subfigure}
\\
\begin{subfigure}[t]{0.245\textwidth}
\centering
\includegraphics[width=.95\textwidth]{Figs_JW/SPECEnsembleMeanBuoyancy_1280.pdf}
\caption{\hfill}
\end{subfigure}
\begin{subfigure}[t]{0.245\textwidth}
\centering
\includegraphics[width=.95\textwidth]{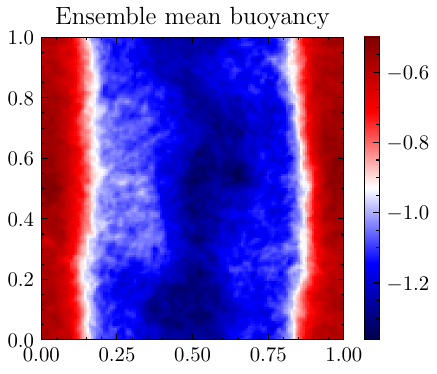}
\caption{\hfill}
\end{subfigure}
\begin{subfigure}[t]{0.245\textwidth}
\centering
\includegraphics[width=.95\textwidth]{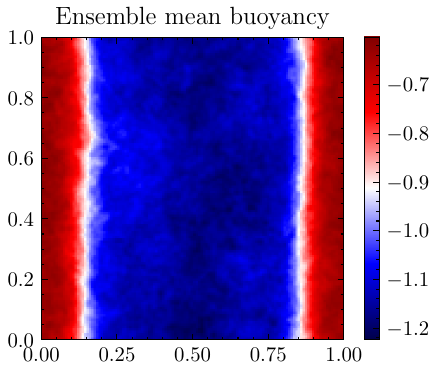}
\caption{\hfill}
\end{subfigure}
\begin{subfigure}[t]{0.245\textwidth}
\centering
\includegraphics[width=.95\textwidth]{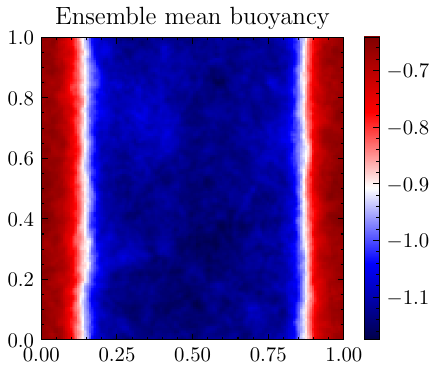}
\caption{\hfill}
\end{subfigure}
\caption{Row 1, the ensemble mean of a 64 member SALT ensemble at times $t \in 2.5 [4]:=\lbrace 2.5,5,7.5,10\rbrace$, during long time behaviour. Row 2, the ensemble mean of a 64 member SPEC ensemble at times $t \in 2.5 [4]$.}
\label{fig:EnsembleMean2}
\end{figure}

The increments used per ensemble solution are computed beforehand such that SPEC or SALT solutions are driven with the same increments $\Delta W_{n,p,E} = \Delta B_{n,p,E} \in \mathbb{R}^{nt \times P\times E}$ and are sampled from the normal distribution. The resulting datasets given by $\lbrace(q^{SALT},b^{SALT})^{n,e}_{i,j} \rbrace_{i,j,n,e\in [n_{x}]\times[n_{x}]\times [n_{t}]\times [E]}$, and $\lbrace(q^{SPEC},b^{SPEC})^{n,e}_{i,j} \rbrace_{i,j,n,e\in [n_{x}]\times[n_{x}]\times [n_{t}]\times [E]}$, (where $(i,j)$ refers to the spatial index of the cell centre, $n$ refers to the time increment, and $e$ refers to the ensemble member) are solutions of the stochastic SALT and SPEC TQG systems, run forward with identical basis and sampled normal increments. We compute the ensemble mean and  ensemble variance of the variable $b$ at time $n$ as follows
\begin{align}
\mathbb{E}_{E}[b] := \frac{1}{E}\sum_{e\in[E]} b_{i,j}^{n,e}, \quad \mathbb{V}_{E}[b] := \mathbb{E}[(b-\mathbb{E}[b])^2] .
\end{align}
The ensemble mean for SALT and SPEC ensembles are plotted in \cref{fig:EnsembleMean} for $t\in 0.625[4]$, during spin up and for $t\in 2.5[4]$ over the full time interval in \cref{fig:EnsembleMean2}. We observe that in the short time behaviour \cref{fig:EnsembleMean} the ensemble mean of SPEC is less blurred than the ensemble mean of SALT, indicating less spread. Where in the long-time behaviour in \cref{fig:EnsembleMean2} the SALT and SPEC have similar ensemble mean behaviour, following the bathymetry variation shown in \cref{fig:bathymetry}. 

\begin{figure}[H]
\centering
\begin{subfigure}[t]{0.245\textwidth}
\centering
\includegraphics[width=.95\textwidth]{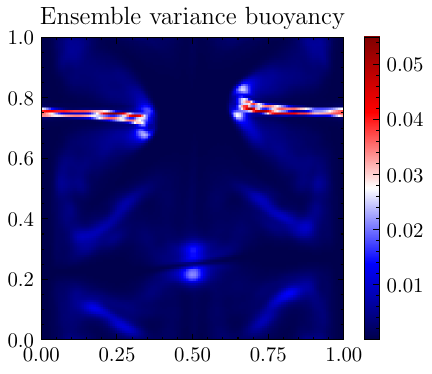}
\caption{\hfill}
\end{subfigure}
\begin{subfigure}[t]{0.245\textwidth}
\centering
\includegraphics[width=.95\textwidth]{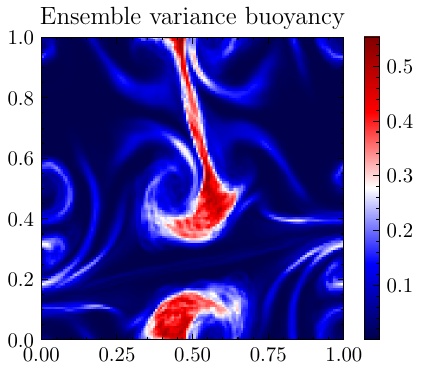}
\caption{\hfill}
\end{subfigure}
\begin{subfigure}[t]{0.245\textwidth}
\centering
\includegraphics[width=.95\textwidth]{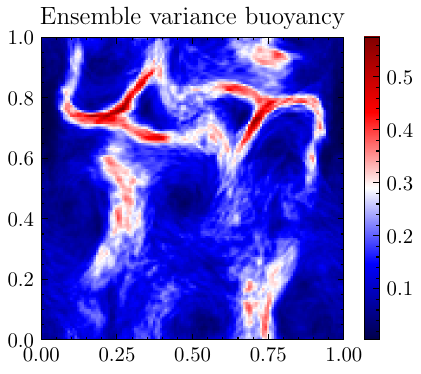}
\caption{\hfill}
\end{subfigure}
\begin{subfigure}[t]{0.245\textwidth}
\centering
\includegraphics[width=.95\textwidth]{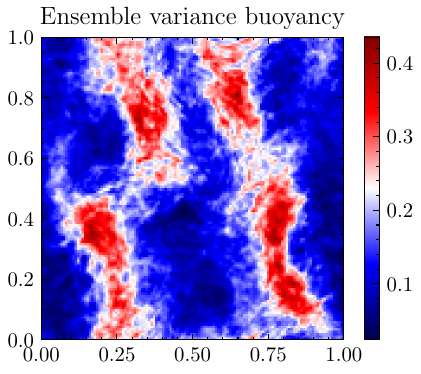}
\caption{\hfill}
\end{subfigure}
\\
\begin{subfigure}[t]{0.245\textwidth}
\centering
\includegraphics[width=.95\textwidth]{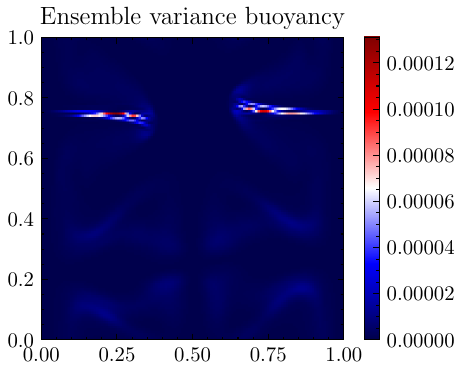}
\caption{\hfill}
\end{subfigure}
\begin{subfigure}[t]{0.245\textwidth}
\centering
\includegraphics[width=.95\textwidth]{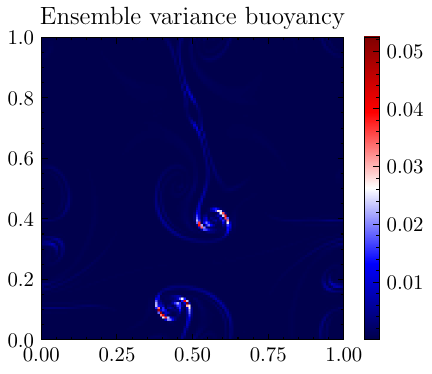}
\caption{\hfill}
\end{subfigure}
\begin{subfigure}[t]{0.245\textwidth}
\centering
\includegraphics[width=.95\textwidth]{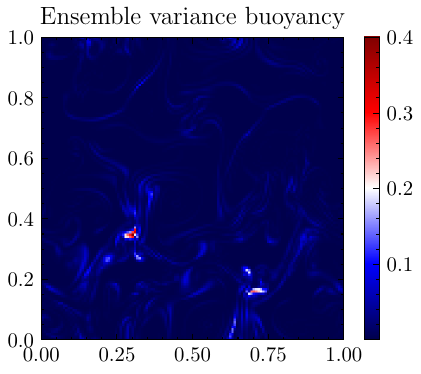}
\caption{\hfill}
\end{subfigure}
\begin{subfigure}[t]{0.245\textwidth}
\centering
\includegraphics[width=.95\textwidth]{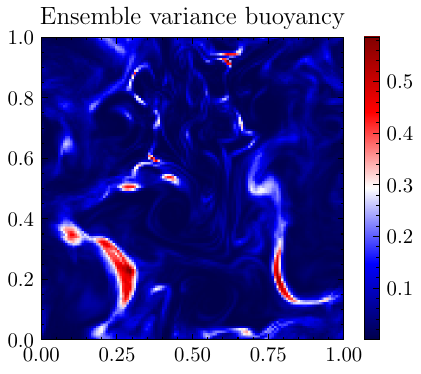}
\caption{\hfill}
\end{subfigure}
\caption{Row 1, the ensemble variance of a 64 member SALT ensemble at times $t\in 0.625[4]:=\lbrace 0.625,1.25,1.875,2.5\rbrace$, during spin-up. Row 2, the ensemble variance of a 64-member SPEC ensemble at times $t\in 0.625[4]$, during spin-up.}
\label{fig:EnsembleVariance}
\end{figure}

\begin{figure}[H]
\centering
\begin{subfigure}[t]{0.245\textwidth}
\centering
\includegraphics[width=.95\textwidth]{Figs_JW/SALTEnsembleVarBuoyancy_1280.pdf}
\caption{\hfill}
\end{subfigure}
\begin{subfigure}[t]{0.245\textwidth}
\centering
\includegraphics[width=.95\textwidth]{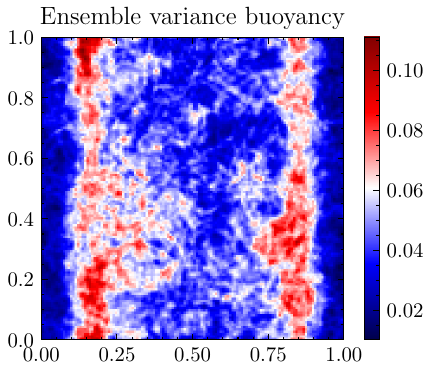}
\caption{\hfill}
\end{subfigure}
\begin{subfigure}[t]{0.245\textwidth}
\centering
\includegraphics[width=.95\textwidth]{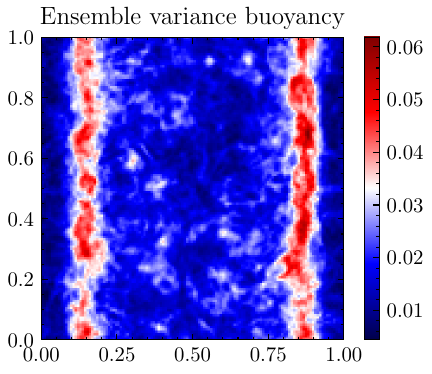}
\caption{\hfill}
\end{subfigure}
\begin{subfigure}[t]{0.245\textwidth}
\centering
\includegraphics[width=.95\textwidth]{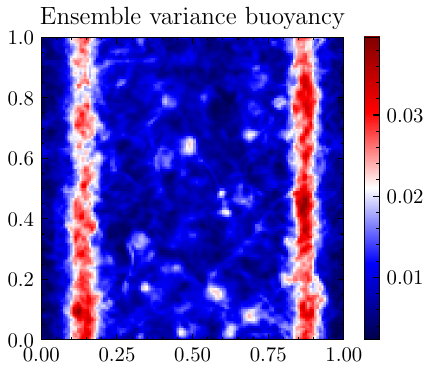}
\caption{\hfill}
\end{subfigure}
\\
\begin{subfigure}[t]{0.245\textwidth}
\centering
\includegraphics[width=.95\textwidth]{Figs_JW/SPECEnsembleVarBuoyancy_1280.pdf}
\caption{\hfill}
\end{subfigure}
\begin{subfigure}[t]{0.245\textwidth}
\centering
\includegraphics[width=.95\textwidth]{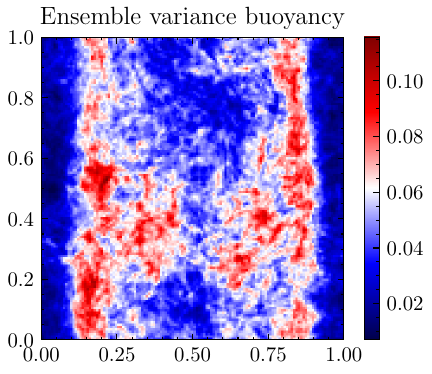}
\caption{\hfill}
\end{subfigure}
\begin{subfigure}[t]{0.245\textwidth}
\centering
\includegraphics[width=.95\textwidth]{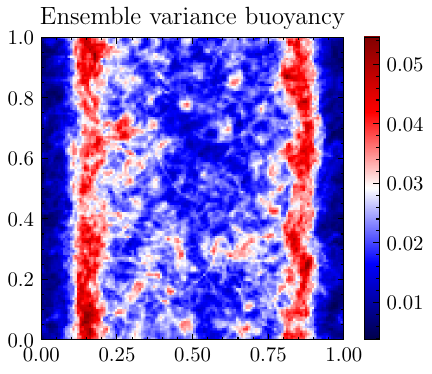}
\caption{\hfill}
\end{subfigure}
\begin{subfigure}[t]{0.245\textwidth}
\centering
\includegraphics[width=.95\textwidth]{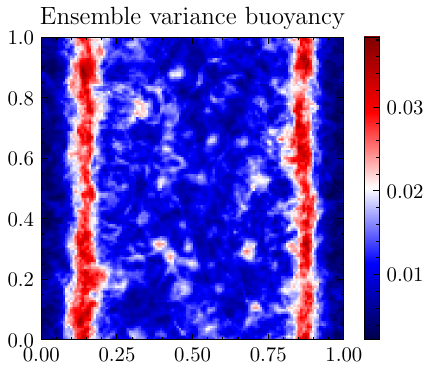}
\caption{\hfill}
\end{subfigure}
\caption{Row 1, the ensemble variance of a 64 member SALT ensemble at times $t\in 2.5[4]:=\lbrace 2.5,5,7.5,10\rbrace$ during long time behaviour. Row 2, the ensemble variance of a 64 member SPEC ensemble at times $t\in 2.5[4]$.}
\label{fig:EnsembleVariance2}
\end{figure}

The ensemble variance for SALT and SPEC ensembles are plotted in \cref{fig:EnsembleVariance} for $t\in 0.625[4]$, during spin up and for $t\in 2.5[4]$ over the full time interval in \cref{fig:EnsembleVariance2}. We observe that in the short time behaviour in \cref{fig:EnsembleVariance} the ensemble variance in buoyancy is several orders of magnitude larger in the SALT ensemble than the SPEC ensemble initially. In \cref{fig:EnsembleVariance} the ensemble variance of buoyancy evolves and the distribution of variance is different in both the SALT and SPEC ensembles. 
However, in the long-time behaviour the ensemble variance observed in \cref{fig:EnsembleVariance2} for both SALT and SPEC appear distributed similarly, and larger variance occurs near the gradients of bathymetry variation \cref{fig:bathymetry}.

\begin{figure}[htp!]
\centering
\begin{subfigure}[t]{0.495\textwidth}
\centering
\includegraphics[width=.95\textwidth]{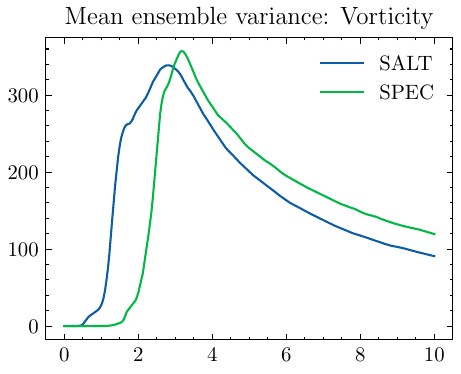}
\caption{\hfill}\label{fig:MEV:q}
\end{subfigure}
\begin{subfigure}[t]{0.495\textwidth}
\centering
\includegraphics[width=.95\textwidth]{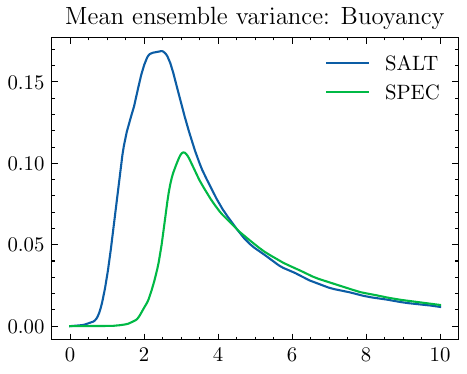}
\caption{\hfill}\label{fig:MEV:b}
\end{subfigure}
\caption{Mean ensemble variance in $(q,b)$ of SALT and SPEC ensembles over time interval $t\in [0,10]$. We observe the ensemble variance is more sensitive to SALT stochastic perturbations, generating a larger spread during the spin-up. However, in the long-time behaviour, TQG buoyancy has a similar mean ensemble variance for both SALT and SPEC perturbations, which decrease with lead time and observe similar spread. }
\label{fig:mean ensemble variance in time}
\end{figure}

We compute the spatial average of the ensemble mean and ensemble variance as follows
\begin{align}
\mathbb{E}_{n_{x},n_{y}}[b] := \frac{1}{n_{x}n_{y}}\sum_{i,j \in [n_x]\times [n_y]} \mathbb{E}_{E}[b], \quad \mathbb{V}_{n_{x},n_{y}}[b]:=\mathbb{E}_{n_{x},n_{y}}[(b-\mathbb{E}_{n_{x},n_{y}}[b])^2].
\end{align}
The spatially averaged ensemble variance for both vorticity and buoyancy is plotted in \cref{fig:mean ensemble variance in time}, against time over $t\in [0,T]$. In \cref{fig:mean ensemble variance in time} for both vorticity and buoyancy variables, on average SALT type perturbations generate more ensemble variance in the initial run-up period, demonstrating an initial larger sensitivity to SALT perturbations. We also observe that ensemble variance increases during spin-up, decreasing as the solution of TQG becomes more stable. Indicating both SALT and SPEC ensembles generate spread sensitive to the sensitivity/stability of the TQG solution. In \cref{fig:mean ensemble variance in time} variance in vorticity appears coupled to variance in buoyancy, as would be expected. In \cref{fig:mean ensemble variance in time} SALT perturbations affect the mean ensemble variance of buoyancy sooner than SPEC perturbations and are larger in magnitude. This can be speculated to occur because SALT noise affects the buoyancy through equation coupling (similar to SPEC) but also by direct stochastic transport \cref{eq:salt-spec-tqg2}.

\begin{figure}[H]
    \centering
\includegraphics[width=1\linewidth]{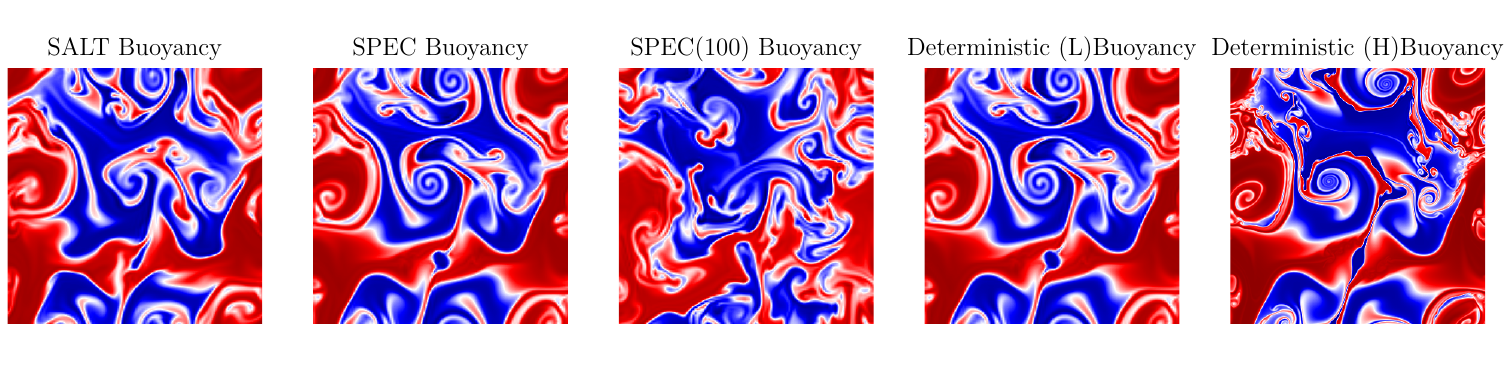}
    \caption{The Buoyancy of a single realisation $(e=1)$ of SALT-TQG, SPEC-TQG, SPEC-TQG with a 100 times larger in magnitude basis of noise, deterministic low-resolution TQG, and deterministic high-resolution TQG all at $t = 1.5625$, with the same driving increments, and basis of noise. During this spin-up phase for this particular realisation, SALT noise perturbed the solution further from the deterministic low-resolution solution than SPEC noise. This is observed in ensemble variance behaviour in \cref{fig:MEV:b} at $t=1.5625$. However, SPEC noise can generate a larger spread with a larger magnitude basis.}
    \label{fig:Snaps.}
\end{figure}

Finally, we refer to \cref{fig:Snaps.} for a single realisation of SALT, SPEC, SPEC(100) using a basis 100 times larger than the other SPEC ensemble, Low-res and High-res dynamics at $t=1.5625$. For an equal basis and driving increments SALT noise perturbed the solution further from the deterministic low resolution than the SPEC noise. However, when the magnitude of the basis is larger, one observes a larger deviation from the deterministic low-resolution solution. 

\section{Overview and outlook}\label{sec: overview and outlook}

In summary, the SPEC approach is consistent with the Euler-Poincar\'e variational principle \cite{holm1998euler} \cref{sec:SPEC+SALT-EP-derivation}; it is compatible with the SALT approach in \cite{holm2015}; and it has a Lie-Poisson structure. SPEC also may be applied to both finite and infinite dimensional problems \cref{sec:examples}, and in this paper SPEC has been shown to produce effects that are similar in magnitude to the effects of SALT but which are qualitatively different in ensemble behaviour for TQG. Namely, by stochastically perturbing potential energy, the ensemble spread sensitivity of the TQG ensemble was altered whilst preserving the Lie-Poisson structure.

In several attempts to calibrate the stochastic basis for noise, an EOF/PCA/SVD decomposition has been employed, see for example \cite{cotter2020particle,cotter2018modelling,woodfield2024stochastic,resseguier2021new,resseguier2020data,crisan2023noise}. Depending on the reference data from which the stochastic basis is calibrated, the resulting stochastic basis will likely generate different spread behaviour, and in some circumstances, may not generate a desirable spread in the solution behaviour amenable to standard data assimilation techniques. Finding a method for generating different types of spread in ensemble behaviour had initially motivated this work and the SPEC approach tested here may be useful to future data assimilation efforts. The TQG system here showed spin-up behaviour that was less sensitive to SPEC stochastic perturbation in bathymetry, whilst still generating spread in long-time dynamics. Whether calibrated basis functions for both SALT and SPEC ensembles would produce greater forecast skill, is of crucial importance, and would likely depend on the type of data one is attempting to model. One can hypothesise that SPEC could be used to model uncertainty in the bathymetry, whilst still requiring a SALT parametrisation to represent uncertainty in the transport velocity, for which additional calibration methodology would need to be developed.

In the modelling context, often TQG/TRSW/RSW/QG models are derived in the multilayer context \cite{beron2021multilayer}, for which stochastically parametrising bathymetry could in principle be interpreted as a stochastic parametrisation of uncertainty in the mixed layer. The application of stochastically perturbing either bathymetry or mixed layer height may also become useful in the context of stochastic parametrisation of multi-layer models. 

\section*{Data availability}
The data that support the findings of this study are available from the corresponding author upon reasonable request.

\section*{Acknowledgements}
We are grateful to our friends, colleagues and collaborators for their advice and encouragement in the matters treated in this paper. 
We especially thank C. Cotter, D. Crisan, R. Hu, E. Luesink, O. D. Street for many insightful discussions of corresponding results similar to the ones derived here for TQG. DH, WP and JW are also grateful for partial support during the present work by European Research Council (ERC) Synergy grant ``Stochastic Transport in Upper Ocean Dynamics" (STUOD) -- DLV-856408.

\bibliography{BIB.bib}
\bibliographystyle{abbrv} 

\appendix

\section{Composition of stochastic transport and stochastic forcing}\label{sec=app}
\subsection{Set up}
Deterministic Lie-Poisson Hamiltonian fluid systems defined the duals of Lie algebras 
comprising semidirect product action of vector fields $\mathfrak{X}(\mathcal{D})$ acting on tensor spaces take the following form in terms of momentum density $m\in \Lambda^1\otimes Den\in \mathfrak{X}^*(\mathcal{D})$, transport velocity vector field  $u\in \mathfrak{X}(\mathcal{D})$ and advected quantities $a\in V^*$ where $V^*$ is 
a tensor space defined in the domain of flow $\mathcal{D}$, \cite{holm1998euler}
\begin{align}
\frac{\partial}{\partial t}
\begin{bmatrix}
m \\ a
\end{bmatrix}
=
-
\begin{bmatrix}
\mathcal{L}_\Box m & \Box \diamond a
\\
\mathcal{L}_\Box a & 0
\end{bmatrix}
\begin{bmatrix}
\delta H / \delta m \\ \delta H / \delta a
\end{bmatrix}
= -
\begin{bmatrix}
\mathcal{L}_{(\delta H / \delta m)} m + (\delta H / \delta a)\diamond a
\\
\mathcal{L}_{(\delta H / \delta m)} a
\end{bmatrix}
.\label{determ-LP}
\end{align}
Here,  the Hamiltonian $H(m,a)$ is a functional of $(m,a)$ and one sums over the advected quantities $a\in V^*$. The symbol $\mathcal{L}_\Box$ denotes the transport Lie derivative which operates on the items in the column vector of variational derivatives of $H(m,a)$.
Likewise $\diamond$ denotes the dual representation of these Lie-derivative operations in 
the following sense of a relation between two different pairings,
\begin{align}
\scp{p\diamond q }{\xi}_{\mathfrak{g}^*\times  \mathfrak{g}}
:= 
\scp{p}{- \mathcal{L}_\xi q}_{V^{*}\times V}
.\end{align}

\medskip
To illustrate the application of this approach, we compute explicit formulae for the diamond operation in the cases that the set of tensor fields $a$ comprises elements with
the following coordinate functions in a Euclidean basis on $\mathbb{R}^3$,
\begin{equation}
a\in\{b,\mathbf{A}\cdot d\mathbf{x},\mathbf{B}\cdot d\mathbf{S},D\,d^3x,
S_{ab}\,dx^a\otimes dx^b\}\,.
\label{Eul-ad-qts}
\end{equation}
These are the tensor fields that typically occur in ideal continuum
dynamics. Thus, for the set of tensor fields $a$ in equation (\ref{Eul-ad-qts}) we
have the following Euclidean $k$-components of
$\frac{\delta H}{\delta a}\diamond a$ \cite{holm1998euler},
\begin{eqnarray} \label{diamond-a-eq}
\left.
\left(    \frac{\delta H}{\delta a}\diamond a
\right)_k
\right.
&=&
-\ \left.
\frac{ \delta H}{ \delta b}\nabla_k b
\ +\ D\nabla_k\frac{ \delta H}{ \delta D}
\right.
\nonumber \\&&
+\ \left.\left(
   -\ \frac{ \delta H}{ \delta \mathbf{A}}\times{\rm curl}\, \mathbf{A}
   +\mathbf{A}\ {\rm div}\,\frac{ \delta H}{ \delta \mathbf{A}}\
   +\mathbf{B}\times{\rm curl}\, \frac{ \delta H}{ \delta \mathbf{B}}
\right)_k
\right.
\nonumber \\ &&
-\ \left.
       \frac{ \delta H}{\delta S_{ab}} \nabla_k S_{ab}
        \ +\ \nabla_a\left(\frac{ \delta H}{\delta S_{ab}}S_{kb}\right)
        \ +\ \nabla_b\left(\frac{ \delta H}{\delta S_{ab}}S_{ka}\right)
\right.\,.
\end{eqnarray}
\subsection{Introducing two types of stochasticity}\label{sec:Introducing two types of stochasticity}
The deterministic Lie-Poisson equations may be made stochastic in two different ways which both preserve the coadjoint semidirect product motion of fluid dynamics with advected quantities. These two approaches involve the formation of stochastic Hamiltonians
which contribute either stochastic transport or stochastic forcing. These two options are called SALT (Stochastic Advection by Lie Transport) and SPEC (Stochastic Potential Energy Coupling). Their simplest implementation augments the deterministic Hamiltonian by adding cylindrical stochasticity which is coupled \emph{linearly} to the momentum $m$ and/or to the advected quantities $a$ as follows,%
\footnote{Nonlinear coupling is allowed, but for clarity of exposition, we deal here only with linear coupling.}
\begin{align}
\mathrm{d}\widetilde{H} := H(m,a)dt + \scp{m}{\sum_{i=1}^N \xi_i(\b x)\circ dW^i_t}
+ \scp{a}{\sum_{j=1}^n \zeta_j(\b x)\circ dB^j_t}
.\label{stoch-lin-couple}
\end{align}
Here $dW^i_t$ and $dB^j_t$ are uncorrelated sets of Brownian motions. The combined SALT and SPEC augmented dynamics will then be expressed as
\begin{align}
\begin{split}
\frac{\partial}{\partial t}
\begin{bmatrix}
m \\ a
\end{bmatrix}
&=
-
\begin{bmatrix}
\mathcal{L}_\Box m & \Box \diamond a
\\
\mathcal{L}_\Box a & 0
\end{bmatrix}
\begin{bmatrix}
\delta (\mathrm{d}\widetilde{H}) / \delta m \\ \delta (\mathrm{d}\widetilde{H}) / \delta a
\end{bmatrix}
\\&=
-
\begin{bmatrix}
\mathcal{L}_\Box m & \Box \diamond a
\\
\mathcal{L}_\Box a & 0
\end{bmatrix}
\begin{bmatrix}
\delta (\mathrm{d}\widetilde{H}) / \delta m  \\ \delta (\mathrm{d}\widetilde{H}) / \delta a
\end{bmatrix}
\\&= -
\begin{bmatrix}
\mathcal{L}_{\delta (\mathrm{d}\widetilde{H}) / \delta m } m
+ (\delta (\mathrm{d}\widetilde{H}) / \delta a)\diamond a
\\
\mathcal{L}_{\delta (\mathrm{d}\widetilde{H}) / \delta a } a
\end{bmatrix}
.\end{split}
\label{stoch-LP}
\end{align}
\subsection{Examples}\label{sec:examples}
\textbf{Example 1: Stochastic Euler fluid.} 
The stochastic Euler fluid equations with SALT and SPEC may be written conveniently as in \eqref{stoch-lin-couple} after using $L^2$ pairing $\scp{}{}$ to express a constrained reduced Hamiltonian with 
1-form density momentum $m=Du^\flat$ as
\begin{align}
 \mathrm{d}\widetilde{H}(m,D) = \frac12\scp{m}{u}dt + \scp{p}{D-1}dt 
 + \scp{m}{\sum_{i=1}^N \xi_i(\b x)\circ dW^i_t}
+ \scp{D}{\sum_{j=1}^n \zeta_j(\b x)\circ dB^j_t}
.\label{StochEul-Ham}
\end{align}
The second term in \eqref{StochEul-Ham} with Lagrange multiplier $p$ imposes volume preservation as $D=1$ for the Jacobian of the Lagrange-to-Euler map, defined as $Dd^3x=d^3x_0$.
The combined SALT and SPEC augmented dynamics will then be expressed as
\begin{align}
\begin{split}
\mathrm{d}
\begin{bmatrix}
m \\ D
\end{bmatrix}
= -
\begin{bmatrix}
\mathcal{L}_{u\,dt + \sum_{i=1}^N \xi_i(\b x)\circ dW^i_t} m 
+ \big(p + \sum_{j=1}^n \zeta_j(\b x)\circ dB^j_t \big)\diamond D
\\
\mathcal{L}_{u\,dt + \sum_{i=1}^N \xi_i(\b x)\circ dW^i_t}  D
\end{bmatrix}
\quad\hbox{with, e.g.,}\quad 
p \diamond D = D \nabla p
\,,
\end{split}
\label{stoch-LPredux}
\end{align}
in which a stochastic part is added to the pressure.
Hence, preservation of the constraint $D=1$ implies the divergence-free condition
for the transport velocity vector field,
\begin{align}
\mathrm{div}\big(u\,dt + \sum_{i=1}^N \xi_i(\b x)\circ dW^i_t\big) = 0
\,.
\label{Eul-LP-div}
\end{align}
Because the passage from the deterministic Euler equations to the SALT and SPEC stochastic equations in \eqref{stoch-LP} leaves the Lie-Poisson operator for the deterministic dynamics invariant, the resulting stochastic Kelvin-Noether circulation theorem still has the same form as in the deterministic case \cite{holm1998euler}. Namely, 
\begin{align}
\mathrm{d}\oint_{u\,dt + \sum_{i=1}^N \xi_i(\b x)\circ dW^i_t}\b u\cdot d\b x
=
\oint_{u\,dt + \sum_{i=1}^N \xi_i(\b x)\circ dW^i_t} d \big(p + \sum_{j=1}^n \zeta_j(\b x)\circ dB^j_t \big) = 0\,.
\label{Eul-circ}
\end{align}
This expression is valid, provided the stochastic processes $dW^i_t$ and $dB^i_t$ are uncorrelated. 
\medskip

\textbf{Example 2: Heavy top.} The Poisson operator  for the stochastic heavy top equations with SALT and SPEC may be written conveniently as in \eqref{stoch-LP} because of the following Lie algebra isomorphisms 
\[
\mathfrak{se}(3)\simeq \mathfrak{so}(3)\times \mathbb{R}^3 \simeq \mathbb{R}^3\times \mathbb{R}^3
\quad\hbox{and}\quad (\mathbb{R}^3)^* \simeq \mathbb{R}^3
\,.\]
These isomorphisms enable the use of $\mathbb{R}^3$ pairing $\scp{\mathbf{\Pi}}{\mathbf{\Gamma}} = \mathbf{\Pi}\cdot\mathbf{\Gamma}$ to express a symmetry-reduced SALT and SPEC stochastic Hamiltonian for the heavy top in terms of body angular momentum $\mathbf{\Pi}\in \mathbb{R}^3$ and unit vector $\mathbf{\Gamma}\in \mathbb{R}^3$ as
\begin{align}
 \mathrm{d}\widetilde{H}(\mathbf{\Pi},\mathbf{\Gamma}) 
 = \frac12\mathbf{\Pi}\cdot \mathbb{I}^{-1}\mathbf{\Pi}dt + mg\boldsymbol{\chi}\cdot \mathbf{\Gamma} dt
 + \mathbf{\Pi}\cdot \sum_{i=1}^N \boldsymbol{\Xi}_i(\b x)\circ dW^i_t
+ \mathbf{\Gamma}\cdot \sum_{j=1}^n \boldsymbol{\Theta}_j(\b x)\circ dB^j_t
.\label{StochHT-Ham}
\end{align}
The combined stochastic SALT and SPEC equations are given by expanding
out the following matrix form,
\begin{align}
\begin{split}
\mathrm{d}
\begin{bmatrix}
\mathbf{\Pi} \\ \mathbf{\Gamma}
\end{bmatrix}
&=
\begin{bmatrix}
\mathbf{\Pi}\times \Box & \mathbf{\Gamma}\times \Box
\\
\mathbf{\Gamma}\times \Box  & 0
\end{bmatrix}
\begin{bmatrix}
\delta (\mathrm{d}\widetilde{H}) / \delta \mathbf{\Pi} 
\\ 
\delta (\mathrm{d}\widetilde{H}) / \delta \mathbf{\Gamma}
\end{bmatrix}
\\&=
\begin{bmatrix}
\mathbf{\Pi}\times \Box & \mathbf{\Gamma}\times \Box
\\
\mathbf{\Gamma}\times \Box  & 0
\end{bmatrix}
\begin{bmatrix}
\mathbb{I}^{-1}\mathbf{\Pi}dt + \sum_{i=1}^N \boldsymbol{\Xi}_i \circ dW^i_t
\\ 
mg\boldsymbol{\chi} dt + \sum_{j=1}^n \boldsymbol{\Theta}_j \circ dB^j_t
\end{bmatrix}
.\end{split}
\label{Stoch-HT-eqns}
\end{align}
Because the Poisson matrix has not been changed in passing from the deterministic heavy top dynamics 
to its combined SALT and SPEC dynamics, the stochastic equations in \eqref{Stoch-HT-eqns} still conserve the Casimir functions $\mathbf{\Pi}\cdot\mathbf{\Gamma}$ and $|\mathbf{\Gamma}|^2$ which are also conserved by the deterministic flow.
\bigskip

\subsection{SALT versus SPEC by themselves for the heavy top}
Heavy top dynamics with SALT alone has been studied in \cite{arnaudon2018noise}. Heavy top dynamics with only SPEC takes the following simple form with only one additional term, which appears as a stochastic torque affecting the body angular momentum $\mathbf{\Pi}$ in the motion equation,
\begin{align}
\begin{split}
\mathrm{d}
\begin{bmatrix}
\mathbf{\Pi} \\ \mathbf{\Gamma}
\end{bmatrix}
&=
\begin{bmatrix}
\mathbf{\Pi}\times \Box & \mathbf{\Gamma}\times \Box
\\
\mathbf{\Gamma}\times \Box  & 0
\end{bmatrix}
\begin{bmatrix}
\mathbb{I}^{-1}\mathbf{\Pi}dt 
\\ 
mg\boldsymbol{\chi} dt 
+ \sum_{j=1}^n \boldsymbol{\Theta}_j \circ dB^j_t
\end{bmatrix}
\\&=
\begin{bmatrix}
\big(\mathbf{\Pi}\times \mathbb{I}^{-1}\mathbf{\Pi} 
+ \mathbf{\Gamma}\times mg\boldsymbol{\chi}\big) dt\,
+ \,\mathbf{\Gamma}\times
\sum_{j=1}^n \boldsymbol{\Theta}_j \circ dB^j_t
\\
\mathbf{\Gamma}\times \mathbb{I}^{-1}\mathbf{\Pi}dt 
\end{bmatrix}
.\end{split}
\label{SPEC-HT-eqns}
\end{align}


\subsection{TQG Transform to canonical Hamiltonian form.}\label{sec:transform to canonical form}

Consider the following Hamiltonian $H = 0.5(||\psi||^{2}_{H^1} - hb)$, where $\psi = (\Delta-1)^{-1}(q-f)$,
\begin{align}
    H_{TQG} &= -\frac{1}{2}\int(q-f)(\Delta-1)^{-1}(q-f)+hb d^2x= -\frac{1}{2}\int(q-f)\psi+hb d^2x,
\end{align}
By taking variational derivatives $(\delta H /\delta q,\delta H /\delta b) = (-\psi , -h/2)$, one can inspect the Hamiltonian structure
\begin{align}
\frac{d}{dt}
\begin{bmatrix}
    q\\
    b
\end{bmatrix}
& = 
\begin{bmatrix}
J(\cdot,q-b)  &J(\cdot , b) \\
J(\cdot,b)&0
\end{bmatrix}
\begin{bmatrix}
\delta H /\delta q\\
\delta H /\delta b
\end{bmatrix}, \label{eq:Hamiltonian_structure_1}
\end{align}
from \cref{eq:bouyancy,eq:vorticity}. By considering the time differential of an arbitrary functional of state variables $F(q,b)$, one obtains the Poisson structure. 
\begin{align}
\frac{d}{dt}F  = \int_{\Omega} \begin{bmatrix}
    \delta F / \delta q \\
    \delta F / \delta b \\
\end{bmatrix}
^{T}
\begin{bmatrix}
    \partial_t q \\
    \partial_t b
\end{bmatrix}
d^2x
= - \int_{\Omega} 
\begin{bmatrix}
\delta F/ \delta q \\ \delta F/ \delta b
\end{bmatrix}
^{T}
\begin{bmatrix}
J(\psi , q-b) + J(h/2 , b ) \\
J(\psi,b)
\end{bmatrix} d^2 x \\
= -\int_{\Omega} 
\begin{bmatrix}
\delta F/ \delta q, \delta F/ \delta b
\end{bmatrix}
\begin{bmatrix}
J( q-b,\cdot) & J(b,\cdot) \\
J(b,\cdot) & 0
\end{bmatrix} 
    \begin{bmatrix}
\delta H /\delta q\\
\delta H /\delta b
\end{bmatrix}
d^2 x:= \lbrace F,H \rbrace_{nc}
\end{align}




Consider the variable $q +b = \tilde{q}$, (consider adding the bottom row of \cref{eq:Hamiltonian_structure_1} to its top row for a possible motivation). So that the Hamiltonian
\begin{align}
H &= -1/2\int (\tilde{q}-b-f)(\Delta-1)^{-1}(\tilde{q}-b-f) +hb d^2 x, \\
&= -1/2\int (\tilde{q}-b-f)\psi +hb d^2 x,\quad \text{where}\quad \psi = (\Delta-1)^{-1}(\tilde{q}-b-f),
\end{align}
has the following variational derivatives 
$(\delta H /\delta \tilde{q} ,\delta H /\delta b) = (-\psi ,\psi -h/2 )$.
Then, by adding rows of the previous Hamiltonian structure, or by direct calculation, one obtains
\begin{align}
\frac{d}{dt}\tilde{q} & = -J(\psi,q)-J(h/2,b)= -J(\psi,\tilde{q}-b)-J(h/2,b)\\
& = J(-\psi,\tilde{q})+J(\psi -h/2,b) = J( \delta H /\delta \tilde{q} ,\tilde{q})+J(\delta H /\delta b,b).
\end{align}
Giving the Hamiltonian structure 
\begin{align}
\frac{d}{dt}
\begin{bmatrix}
    \tilde{q}\\
    b
\end{bmatrix}
& = 
\begin{bmatrix}
J(\cdot,\tilde{q}) &J(\cdot , b) \\
J(\cdot,b) &0
\end{bmatrix}
\begin{bmatrix}
\delta H /\delta \tilde{q}\\
\delta H /\delta b
\end{bmatrix} ,\label{eq:Hamiltonian_structure 2.}
\end{align}
in the new variables $(\tilde{q},b) = (q+b,b)$. 
Which on considering a time derivative of an arbitrary functional of (modified) state variables $F(\tilde{q},b)$, as before, one obtains the Lie--Poisson bracket
\begin{align}
\frac{dF}{dt}= \int_{\Omega} 
\begin{bmatrix}
\delta F/ \delta  \tilde{q} \\ \delta F/ \delta b
\end{bmatrix}
^{T}
\begin{bmatrix}
J( \cdot,\tilde{q}), &J(\cdot,b) \\
J(\cdot,b) ,& 0
\end{bmatrix} 
    \begin{bmatrix}
\delta H /\delta \tilde{q}\\
\delta H /\delta b
\end{bmatrix}
d^2 x:= \lbrace F,H \rbrace_{can}.
\end{align}

In the variables $(\tilde{q},b)$, the SALT-SPEC-TQG equations may then be expressed as
\begin{align}
d \tilde{q} + (\b u dt+\sum_{i}\b \xi_{i} \circ dW^i)\cdot \nabla \tilde{q} + (\b u dt+\sum_{i}\b \xi_i \circ dW^i+\b u_h dt+\sum_{i}\b \eta_i \circ dB^i)\cdot \nabla b=0 \label{eq:salt-spec-tqg2-can1},\\
db + (\b u dt+\sum_{i}\b \xi_{i}(x)\circ dW^{i})\cdot \nabla b =0. \label{eq:salt-spec-tqg2-can2}
\end{align}

\subsection{SPEC+SALT-EP-derivation}\label{sec:SPEC+SALT-EP-derivation}

It may be worth informally remarking on the connection between the present work and the Euler-Poincare theory developed in \cite{holm1998euler}, in the specific application in continuum dynamics. The following constrained stochastic variational principle,
\begin{align}
0=\delta\int_{t_{1}}^{t_{2}} \ell(u,a) dt + \underbrace{ \sum_{i}\Xi_i(a) \circ dB^i}_{SPEC},\quad \delta a = -\mathcal{L}_{v} a \in V^*, \quad \delta u = d v-\underbrace{\mathcal{L}_{u dt+ \sum_{i}\xi_i \circ dW^i}}_{SALT} v \in \mathfrak{X}(M).
\end{align} 
perturbs the Lagrangian $\ell(u,a)$ by the family of stochastic potential energies $\lbrace\Xi_i(a)\rbrace_{i\in [m]}$ as presented in \cite{street2023semimargingale}, and at the level of the Lie algebra by $\lbrace\xi_i(a)\rbrace_{i\in [m]}\in \mathfrak{X}(M)=\mathfrak{g}$ as presented in \cite{holm2015}, resulting in 
\begin{align}
0=\int_{t_{1}}^{t_{2}}\left\langle \frac{\delta \ell}{\delta u}, d v - \mathcal{L}_{udt + \sum_{i}\xi_i \circ dW^i} v \right\rangle_{\mathfrak{X}(M)} + \left\langle \frac{\delta \ell}{\delta a} + \sum_{i}\frac{\delta \Xi_{i}
}{\delta a}\circ dB^{i}, -\mathcal{L}_{v}a\right\rangle_{V^*}.
\end{align}
Upon identifying the Lie derivative transpose is the negative Lie derivative $\mathcal{L}_{v}^T = -\mathcal{L}_{v}$, the integration is Stratonovich and using $\langle b, -\mathcal{L}_{v}a \rangle_{V^*} = \langle b\diamond a,v\rangle_{\mathfrak{X}(M)}$, (in direct analogy with the deterministic theory presented in \cite{holm1998euler})
integration by parts gives SALT-SPEC-EP(P) equations
\begin{align}
d \frac{\delta \ell}{\delta u} + \mathcal{L}_{u dt + \sum_{i}\xi_i \circ dW^i} \frac{\delta \ell}{\delta u} + \left(\frac{\delta \ell}{\delta a} dt + \sum_{i}\frac{\delta \Xi_i}{\delta a}\circ dB^i \right)\diamond a = 0,\label{eq:SALTSPECEPP1}\\
d a + \mathcal{L}_{u dt + \sum_{i}\xi_i \circ dW^i}a = 0. \label{eq:SALTSPECEPP2}
\end{align}
These equations can be understood as Euler-Poincare (Poisson) equations on a semi-direct product of Lie algebra and vector space, in the symmetry-breaking sense \cite{holm1998euler,holm2015}, elaborated on more formally in the stochastic setting in \cite{street2023semimargingale}. Heuristically, \cref{eq:SALTSPECEPP1} is interpreted weakly in space, evolving a one-form density pathwise in time and \cref{eq:SALTSPECEPP2} is an enforced relation regarding the stochastic pushforward representation on advected quantities \cite{de2020implications,holm1998euler}. Upon taking variational derivatives (w.r.t. $a,m,u$) of the stochastic ``Legendre" transform,
\begin{align}
d\tilde{h}(m,a) &= \langle m,u dt + \sum_{i}\xi_{i}\circ dW^i \rangle_{\mathfrak{g}^*} -\ell(u,a)dt -\sum_{i} \Xi_i(a)\circ dB^{i}, \\
& = h(m,a)dt + \langle m, \sum_{i} \xi_i \circ dW^{i} \rangle_{\mathfrak{g}^*}  - \sum_{i} \Xi_i(a) \circ dB^i \,,
\end{align}
one obtains the following variational expressions,
\begin{align}
\frac{\delta d\tilde{h}}{\delta m
} = \frac{\delta h}{\delta m}dt+\sum_{i}\xi_i\circ dW^{i}= udt+\sum_{i}\xi_{i} \circ dW^{i} \in \mathfrak{X}(M),\\ 
\frac{\delta d\tilde{h}}{\delta a
} =\frac{\delta h}{\delta a}dt -\sum_{i}\frac{\delta \Xi_i}{\delta a}\circ dB^i= -\frac{\delta \ell}{\delta a}dt-\sum_{i}\frac{\delta \Xi_i}{\delta a}\circ dB^i \in V, \\
\frac{\delta \ell}{\delta u} = m \in \mathfrak{X}(M)^*.
\end{align}
Upon substitution into the SALT-SPEC-Euler-Poincare equation \cref{eq:SALTSPECEPP} these expressions yield 
\begin{align}
d m &= - \mathcal{L}_{\frac{\delta h}{\delta m}dt+ \sum_i \xi_i \circ dW^{i}} m - \left(-\frac{\delta h}{\delta a}dt + \frac{\delta \Xi_{i}}{\delta a}\circ dB^i \right)\diamond a,\\
d a &= - \mathcal{L}_{\frac{\delta h}{\delta m}dt + \xi_i \circ dW^{i}} a, \label{eq:SALTSPECEPP}
\end{align}
which upon identifying $\frac{\delta \Xi_{i} }{\delta a } = -\,\zeta_i$ yields the Lie-Poisson form \cref{stoch-LP}. Many systems possess a Hamiltonian Lie--Poisson structure, even without an Euler--Poincar\'e variational principle, such as the TQG equations discussed in this paper. Both SALT-SPEC stochastic perturbations preserve Casimirs as a consequence of this structure, more specifically the Lie Poisson bracket
$$dF =
-\int 
[\frac{\delta F}{\delta m},   \frac{\delta F}{\delta a}
]
\begin{bmatrix}
\mathcal{L}_{(\cdot) }m, & (\cdot)\diamond a\\
 \mathcal{L}_{(\cdot)} a, & 0
\end{bmatrix}
\begin{bmatrix}
\frac{\delta h}{\delta m}dt + \sum_{i}\xi_i \circ dW^{i} \\
\frac{\delta h}{\delta a} dt + \sum_i \frac{\delta \Xi_i}{\delta a} \circ dB^i
\end{bmatrix}d^n x$$
is unchanged; so it contains the same degeneracy (null eigenvectors) as the deterministic system.

\end{document}